\documentclass[12pt]{iopart}

\usepackage{graphicx}
\usepackage{bm}
\usepackage{color}
\usepackage{iopams}

\newcommand{\leftg}{\langle \phi_0 |}
\newcommand{\rightg}{| \phi_0 \rangle}

\begin{document}

\title{Optimizing relativistic energy density functionals: covariance analysis}
\author{T. Nik\v si\' c$^{1}$, N. Paar$^{1}$, P.-G. Reinhard$^{2}$,
D. Vretenar$^{1}$}

\address{$^{1}$Physics Department, Faculty of Science, University of Zagreb,
Zagreb/Croatia}
\address{$^{2}$Institut f\"ur Theoretische Physik, 
Universit\"at Erlangen/N\"urnberg, Erlangen/Germany}
\date{\today}

\begin{abstract}
The stability of model parameters for a class of relativistic energy density functionals, 
characterized by contact (point-coupling) effective inter-nucleon interactions and 
density-dependent coupling parameters, is 
analyzed using methods of statistical analysis. A set of pseudo-observables in 
infinite and semi-infinite nuclear matter is used to define a quality measure
$\chi^2$ for subsequent analysis.
We calculate uncertainties of model parameters and correlation coefficients between parameters, 
and determine the eigenvectors and eigenvalues of the matrix of second derivatives of $\chi^2$ at the minimum. 
This allows to examine the stability of the density functional in nuclear matter, and to deduce weakly 
and strongly constrained combinations of parameters. In addition, we also compute 
uncertainties of observables that are not included in the calculation of $\chi^2$: binding energy 
of asymmetric nuclear matter, surface thickness of semi-infinite nuclear matter, binding energies and 
charge radii of finite nuclei.
\end{abstract}

\maketitle

\section{\label{secI}Introduction}

This special isssue is devoted to model analysis in connection with
least-squares ($\chi^2$) fits and subsequent statistical analysis
\cite{Bra97a}. We will use these 
techniques to explore the details of a particular class of the
relativistic mean field model (RMF), namely the density-dependent
point-coupling model DD-PC1 \cite{Nik08}. The RMF, similar as its
non-relativistic counterparts Skyrme-Hartree-Fock (SHF) and Gogny force
\cite{Ben03aR}, belongs to the family of self-consistent nuclear
models based either on effective interactions or nuclear energy density functionals.
These models necessitate calibration of the model parameters
to empirical data, usually bulk properties of nuclear ground
states. In the early days of RMF models coupling and mass parameters were 
determined basically by an intuitive search \cite{Ser86aR}. A first systematic
calibration using a straightforward $\chi^2$-fit was performed in Ref.
\cite{Rei86a}. The fitting protocol and data pool were similar to those applied within
the SHF framework \cite{Fri86a}. Since then RMF parameterisations have been significantly 
improved by including more and more data in the data pool (for
reviews see Refs. \cite{Rei89aR,Rin96aR,Vre05aR,Meng06}). The traditional RMF approach 
was based on nucleon-meson couplings, that is, on finite-range effective interactions. 
An alternative is the point-coupling (PC) RMF which is closer in spirit
to the non-relativistic SHF approach and allows a simpler treatment of 
exchange terms.  It was first introduced in \cite{Nik92a} and, by performing 
$\chi^2$ fits of model parameters \cite{Bue02a}, its performance has been 
improved to match the level of traditional RMF models. A significant 
step forward in the flexibility of RMF models was the introduction of 
an explicit density dependence of coupling parameters  
\cite{Typ99a}. This idea was later used also in the PC-RMF approach, 
and an empirical ansatz for the density dependence of the nucleon-vertex 
parameters was employed in the formulation of the functional 
DD-PC1 \cite{Nik08}, which will be the basis of the
present study.

Although one could consider different classes of RMF models for
the statistical analysis \cite{Pie12,Rei13a}, in this work we explore
the intrinsic structure of one particular model, i.e. the functional DD-PC1. 
We follow the strategy of the model
study \cite{Fat11} that used pseudo-data related to infinite nuclear matter to
generate a $\chi^2$ measure, which was then analyzed according to the
rules of statistical analysis. This procedure provides theoretical uncertainties for both 
model parameters and predicted observables. The results of the present investigation are 
general and can also be applied to other classes of the RMF models because, by
analyzing the inter-dependencies between the couplings, one is able to determine strongly
and weakly constrained combinations of model parameters and estimate their impact
in extrapolating uncertainties.

\section{\label{secII}Covariance analysis}

Following the notation of Ref.~\cite{Rei10}, we denote a point in an
$F$-dimensional parameter space by $\mathbf{p}=\{p_1,\dots,p_F\}$ 
and, therefore, each value of $\mathbf{p}$ corresponds to a particular model. 
Model parameters are usually adjusted to
ground-state properties of a selected set of finite nuclei, and
  occasionally also to bulk empirical properties of infinite nuclear matter. A
  calibration of model parameters starts with defining a quality
  measure:
\begin{equation}
  \chi^2(\mathbf{p}) 
  =
  \sum_{n=1}^N{\left(\frac{\mathcal{O}_n^{(th)}(\mathbf{p})-\mathcal{O}_n^{(exp)}}
                         {\Delta \mathcal{O}_n} \right)^2},
\end{equation}
where $N$ is the number of observables
$\hat{\mathcal{O}}_n$ considered in the analysis, 
{\em (th)} and {\em (exp)} denote theoretical and experimental values, respectively. 
Every observable is weighted by the
  inverse of $\Delta \mathcal{O}_n$ which, in fits to experimental results, is typically associated 
  with the accuracy of the measurement. In
  calibrating a model one often uses an ``adopted error'' which 
  is supposed to include all sources of uncertainty and is adjusted in
  such a way that $\chi^2(\mathbf{p}_0)\approx N-F$ \cite{Bra97a,Dob14a}.
The optimal set $\mathbf{p}_0$, that is, the ``best model'' corresponds to the minimum of 
$\chi^2$ on the
multidimensional parameter surface, and this implies that all first derivatives of
the function $\chi^2$ vanish at $\mathbf{p}_0$:
\begin{equation}
\left.\frac{\partial \chi^2(\mathbf{p}) }{\partial p_i}\right|_{\mathbf{p}=\mathbf{p}_0}
=0,\quad \forall \; i=1,\dots ,F.
\end{equation}
Moreover, the symmetric $F\times F$ matrix of second derivatives
${\partial^2 \chi^2 }/{(\partial p_i \partial p_j)}$
has to be positive-definite at  $\mathbf{p}_0$.
To analyze the deviation of $\chi^2$ from its minimum value, it is convenient to define dimensionless
parameters
\begin{equation}
  x_i 
  =
  \frac{(\mathbf{p}-\mathbf{p}_0)_i}{(\mathbf{p}_0)_i} 
  .
\label{eq:scale}
\end{equation}
The minimum is then determined by $\mathbf{x}=0$. In the vicinity of
the minimum $\chi^2$ can be represented by a Taylor series expansion. The lowest-order
(quadratic) deviation of $\chi^2$ reads
\begin{eqnarray}
  \Delta \chi^2(\mathbf{x})
  &=& 
  \chi^2(\mathbf{p}) - \chi^2(\mathbf{p}_0) 
  =
  \mathbf{x}^T \hat{\mathcal{M}} \mathbf{x}
  ,
\\ 
  \mathcal{M}_{ij} 
  &=&
   \frac{1}{2} 
   \left. \frac{\partial^2 \chi^2 }{\partial x_i \partial x_j} \right|_{\mathbf{x}=0}
  = 
  \frac{1}{2} \left( \mathbf{p}_0\right)_i  \left(
   \mathbf{p}_0\right)_j 
   \partial_i \partial_j \chi^2(\mathbf{p}_0)
   .  \label{M}
\end{eqnarray}
The curvature matrix $\hat{\mathcal{M}}$ is symmetric and can be
diagonalized by an orthogonal transformation
$\hat{\mathcal{M}}=\hat{\mathcal{A}}\hat{\mathcal{D}}\hat{\mathcal{A}}^T$,
where $\hat{\mathcal{A}}$ denotes the orthogonal matrix with columns
corresponding to normalised eigenvectors of $\hat{\mathcal{M}}$, and 
the diagonal matrix $\hat{\mathcal{D}}$ contains the 
eigenvalues of $\hat{\mathcal{M}}$.  The deviation of 
$\chi^2$ from its minimum value can therefore be expressed as \cite{Fat11}
\begin{equation}
  \Delta \chi^2(\mathbf{x}) 
  = 
  \mathbf{x}^T \left( \mathcal{A}\mathcal{D}\mathcal{A}^T  \right)
  \mathbf{x} 
  = 
  \mathbf{\xi}^T \mathcal{D} \mathbf{\xi} = \sum_{i=1}^F{\lambda_i
    \xi_i^2} 
.
\end{equation}
The transformed vectors $\mathbf{\xi}=\hat{\mathcal{A}}^T\mathbf{x}$ define 
the principal axes on the $F$-dimensional surface in parameter
space.  Soft directions are characterized by small eigenvalues
$\lambda_i$, that is, there is very little deterioration in the function 
$\chi^2$ as one moves along a direction defined by the eigenvector 
that corresponds to a small eigenvalue of $\hat{\mathcal{M}}$. 
This implies that the 
corresponding linear combinations of model parameters are badly
constrained by the observables included in the $\chi^2$ adjustment.  
On the other hand, stiff directions are characterized
by large eigenvalues $\lambda_i$, that is, the function $\chi^2$
increases rapidly along these directions and the corresponding linear
combinations of parameters are tightly constrained by the observables 
that determine the $\chi^2$ measure. A survey
of the eigenvalues $\lambda_i$ and eigenvectors $\xi_i$ of $\hat{\mathcal{M}}$
for the relativistic functional DD-PC1 presents the major part of 
the analysis performed in this work.

Another important concept in statistical analysis is the covariance
between two observables $A$ and $B$~\cite{Bra97a,Rei10,Dob14a}:

\begin{equation}
 \textnormal{cov}(A,B) 
  = 
  \sum_{i,j=1}^F{\frac{\partial A}{\partial x_i}(\mathcal{M}^{-1})_{ij}
              \frac{\partial B}{\partial x_j} } 
  = 
  \sum_{i=1}^F{\frac{\partial A}{\partial \xi_i} \lambda_i^{-1} 
             \frac{\partial B}{\partial \xi_i}} \;, 
\label{eq:mixvariances}
\end{equation}
from which one defines the correlation coefficient
\begin{equation}
  \rho(A,B) 
  = \frac{\textnormal{cov}(A,B)}{\sqrt{\textnormal{var}(A) \textnormal{var}(B)}} \;,
\end{equation}
and where the variance of an observable is simply: $\textnormal{var}(A) = \textnormal{cov}(A,A)$. 
The observables $A$ and $B$ are fully correlated if $ \rho(A,B) = 1$, 
anti-correlated if $ \rho(A,B) = -1$, and independent if $ \rho(A,B) = 0$.

For the calculation of covariances with
 Eq.~(\ref{eq:mixvariances}), one has to compute derivatives of
observables with respect to model parameters. Here we use the
Richardson extrapolation method~\cite{Ric27} for the numerical 
calculation of derivatives and define the $m$-th order expression 
\begin{equation}
G_m(h) = \frac{4^m G_{m-1}(h/2)-G_{m-1}(h)}{4^m-1},\quad m=1,2,\dots,
\label{eq:interpol1}
\end{equation}
where
\begin{equation}
G_0(h) = \frac{1}{2h}\left[ f(a+h) -f(a-h)   \right] 
\label{eq:interpol2}
\end{equation}
is a simple first-order finite difference, and thus 
$f^\prime (a) - G_m(h) = O(h^{2(m+1)})$. 
As in the recent analysis of the propagation of uncertainties in Skyrme 
energy density functionals of Ref.~\cite{Gao13}, here we extrapolate up to
$m=2$ in the computation of derivatives with respect to model parameters.

\section{\label{secIII}  The Relativistic Density Functional DD-PC1}

The basic building blocks of a relativistic nuclear energy density
functional are the densities and currents bilinear in the Dirac spinor
field $\psi$ of the nucleon: $\bar{\psi}\mathcal{O}_\tau \Gamma \psi$,
with $\mathcal{O}_\tau \in \{1,\tau_i\}$ and $\Gamma \in
\{1,\gamma_\mu,\gamma_5,\gamma_5\gamma_\mu,\sigma_{\mu\nu}\}$.  Here
$\tau_i$ are the isospin Pauli matrices and $\Gamma$ generically
denotes the Dirac matrices.  A general covariant Lagrangian can
be written as a power series in the currents
$\bar{\psi}\mathcal{O}_\tau\Gamma\psi$ and their derivatives. The
point-coupling functional or point-coupling relativistic mean-field (PC-RMF) 
model is structurally particularly simple as it involves only contact 
couplings between these currents.  Actually we take into account the
following channels: 
isoscalar-scalar $(\bar\psi\psi)^2$, 
isoscalar-vector $(\bar\psi\gamma_\mu\psi)(\bar\psi\gamma^\mu\psi)$, and
isovector-vector $(\bar\psi\vec\tau\gamma_\mu\psi)
\cdot(\bar\psi\vec\tau\gamma^\mu\psi)$.
The Lagrangian of the PC-RMF model contains four-fermion
(contact) interaction terms: 
\begin{eqnarray}
\label{Lagrangian}
\mathcal{L} &=& \bar{\psi} (i\gamma \cdot \partial -m)\psi \nonumber \\
     &-& \frac{1}{2}\alpha_s(\hat{\rho}_v)(\bar{\psi}\psi)(\bar{\psi}\psi)
       - \frac{1}{2}\alpha_v(\hat{\rho}_v)(\bar{\psi}\gamma^\mu\psi)(\bar{\psi}\gamma_\mu\psi)
 \nonumber \\
     &-& \frac{1}{2}\alpha_{tv}(\hat{\rho_v})(\bar{\psi}\vec{\tau}\gamma^\mu\psi)
                                                                 (\bar{\psi}\vec{\tau}\gamma_\mu\psi)
 \nonumber \\
    &-&\frac{1}{2} \delta_s (\partial_\nu \bar{\psi}\psi)  (\partial^\nu \bar{\psi}\psi)
         -e\bar{\psi}\gamma \cdot A \frac{(1-\tau_3)}{2}\psi\;.
\end{eqnarray}
The last term therein defines the coupling of the protons to
  the electromagnetic four-potential.  The derivative term in
  Eq.~(\ref{Lagrangian}) accounts for next-order effects from a 
  density-matrix expansion of finite-range and correlation effects
  \cite{Neg72a,Rei94aR}. Although one could include a derivative
term in each spin-isospin channel, in practice ground-state data can
constrain only a single derivative term.  In particular, DD-PC1 implements this term in
the isoscalar-scalar channel.
The strength parameters $\alpha_c$ of the interaction terms in
Eq.~(\ref{Lagrangian}) are density-dependent functionals 
  of $\sqrt{j^\mu j_\mu}$, with the nucleon 4-current:
  $j^\mu = \bar{\psi} \gamma^\mu \psi$.  However, at low velocities
  relevant for the present investigation, the parameters $\alpha_c$ depend only on the
  baryon density $\hat{\rho}_v=\psi^\dagger\psi$. The single-nucleon
  Dirac equation, that is, the relativistic analogue of the Kohn-Sham equation
  \cite{Dre90}, is obtained from the variation of the Lagrangian with
respect to $\bar{\psi}$. This yields:
\begin{equation}
\left[ \gamma_\mu(i\partial^\mu - \Sigma^\mu -\Sigma_R^\mu) - (m+\Sigma_S)\right]\psi = 0\;,
\label{Dirac-eq}
\end{equation}
with the nucleon self-energies $\Sigma$ defined by the following relations:
\begin{eqnarray}
\label{sigma_v}
\Sigma^\mu &=& \alpha_V(\rho_v) j^\mu + e  \frac{(1-\tau_3)}{2} A^\mu\\
\label{sigma_r}
\Sigma_R^\mu &=& \frac{1}{2}\frac{j^\mu}{\rho_v}
            \left\{ \frac{\partial \alpha_s}{\partial \rho}\rho_s^2
         +\frac{\partial \alpha_v}{\partial \rho}j_\mu j^\mu
         + \frac{\partial \alpha_{tv}}{\partial \rho}\vec{j}_\mu \vec{j}^\mu
          \right\}\\
\label{sigma_s}
\Sigma_S &=& \alpha_s(\rho_v)\rho_s - \delta_s \Box \rho_s \\
\label{sigma_tv}
\Sigma_{TV}^\mu &=&  \alpha_{tv}(\rho_v)\vec{j}^\mu \;.
\end{eqnarray}
In addition to contributions from the isoscalar-vector four-fermion
interaction and the electromagnetic interaction, the isoscalar-vector
self-energy includes the ``rearrangement'' terms
$\Sigma_R^\mu$ that arise from the variation of the vertex functionals
$\alpha_s$, $\alpha_v$, and $\alpha_{tv}$ with respect to the nucleon
fields in the vector density operator $\hat{\rho}_v$.
The importance of self-energies $\Sigma$ becomes more apparent in the 
  non-relativistic limit \cite{Rei89aR}. Of particular interest is the 
  interplay between the scalar $\Sigma_S$ and the vector component
  $\Sigma^0$. The non-relativistic local mean-field potential is
  determined by the sum $\Sigma_S+\Sigma^0$, which is relatively small because these 
  self-energies have 
  opposite signs. The difference $\Sigma_S-\Sigma^0$, on the other
  hand, is large and explains the comparatively large energy spacings between spin-orbit
  partner states \cite{Due56a}.  Thus the sum and difference of the isoscalar-scalar and 
  isoscalar-vector self-energies relate to very different physical effects.

At the mean-field level the nuclear ground state $\rightg$ is
represented by the self-consistent solution of the system of equations
(\ref{Dirac-eq}) -- (\ref{sigma_tv}), with the ground-state isoscalar and isovector
four-currents and scalar density defined as expectation values:
\begin{eqnarray}
\label{den1}
j_\mu & = \leftg \bar{\psi} \gamma_\mu \psi \rightg =
& \sum_{k=1}^N v_{k}^{2}~\bar{\psi}_k \gamma_\mu \psi_k \; ,\\
\label{den2}
\vec{j}_\mu & =
\leftg \bar{\psi} \gamma_\mu \vec{\tau} \psi \rightg =
& \sum_{k=1}^N v_{k}^{2}~\bar{\psi}_k \gamma_\mu \vec{\tau} \psi_k \; ,\\
\label{den3}
\rho_s & = \leftg \bar{\psi} \psi \rightg = &
\sum_{k=1}^N v_{k}^{2}~\bar{\psi}_k \psi_k \; ,
\end{eqnarray}
where $\psi_k$ are Dirac spinors, and the sum runs over occupied
positive-energy single-nucleon orbitals, including the corresponding
occupation factors $v_{k}^{2}$.  The single-nucleon Dirac equations
are solved self-consistently in the {\em ``no-sea''} approximation, that
omits explicit contributions of negative-energy solutions of the
relativistic equations to densities and currents
\cite{Ser86aR,Rei89aR,Rin96aR}.

In a phenomenological construction of a relativistic energy density
functional one starts from an assumed ansatz for the medium dependence
of the mean-field nucleon self-energies, and adjusts the free
parameters directly to ground-state data on finite nuclei. Guided by
the microscopic density dependence of the vector and scalar
self-energies, the following practical ansatz for the functional form
of the couplings was adopted in Ref.~\cite{Nik08}:
\begin{eqnarray}
\alpha_s(\rho)&=& a_s + (b_s + c_s x)e^{-d_s x},\nonumber\\
\alpha_v(\rho)&=& a_v +  b_v e^{-d_v x},\label{parameters} \\
\alpha_{tv}(\rho)&=& b_{tv} e^{-d_{tv} x},\nonumber
\end{eqnarray}
with $x=\rho/\rho_\mathrm{sat}$, where $\rho_\mathrm{sat}$ denotes the nucleon
density at saturation in symmetric nuclear matter.  The set of 10
strength parameters was adjusted in a multistep parameter fit
exclusively to the experimental masses of 64 axially deformed nuclei
in the mass regions $A\approx 150-180$ and $A\approx 230-250$. The
resulting functional DD-PC1 \cite{Nik08} has been further tested in
calculations of binding energies, charge radii, deformation
parameters, neutron skin thickness, and excitation energies of giant
monopole and dipole resonances.  The parameters of DD-PC1 are given in
Table~\ref{DD-PC1}.  The nuclear matter equation of state that 
corresponds to DD-PC1 is characterized by the following properties
at the saturation point: nucleon density
$\rho_\mathrm{sat}=0.152~\textnormal{fm}^{-3}$, volume energy $a_v=-16.06$
MeV, surface energy $a_s=17.498$ MeV, symmetry energy $a_4 = 33$ MeV,
and the nuclear matter compression modulus $K_{nm} = 230$ MeV
(see also table \ref{Tab:inf-nuc-mat}).

\begin {table}[tbp]
\begin {center}
\caption{Parameters of the relativistic energy density functional DD-PC1
(cf. Eq.~(\ref{parameters})). The value of the nucleon mass is $m=939$ MeV.}
\bigskip
\begin {tabular}{cc}
\hline
\hline
 {\sc parameter} &
 \\ \hline
$a_s$ (fm$^2$)    & $-10.0462$  \\
$b_s$ (fm$^2$)   & $-9.1504$ \\
$c_s$ (fm$^2$)    & $-6.4273$  \\
$d_s$        & $1.3724$   \\ \hline
$a_v$  (fm$^2$)   & $5.9195$  \\
$b_v$  (fm$^2$)  & $8.8637$ \\
$d_v$        & $0.6584$    \\ \hline
$b_{tv}$ (fm$^2$) & $1.8360$ \\
$d_{tv}$ & $0.6403$\\ \hline
$\delta_s$ (fm$^4$) & $-0.8149$ \\
\hline
\end{tabular}
\label{DD-PC1}
\end{center}
\end{table}

Although we could analyze correlations between the individual
parameters $a_i$, $b_i$, $c_i$ and $d_i$ in Eq.~(\ref{parameters}), we choose to 
examine correlations between the lowest-order terms in a Taylor expansion of the 
density-dependent coupling functions around the 
saturation point: $\alpha_i(\rho_\mathrm{sat})$,
$\alpha_i^\prime(\rho_\mathrm{sat})$ and $\alpha_i^{\prime
  \prime}(\rho_\mathrm{sat})$, because these quantities directly  
  determine the expressions
for the binding energy, pressure and compressibility of nuclear
matter at saturation.  Also, such an analysis could lead to more general
conclusions that can be related to other density functionals, rather than 
just the one defined by the couplings in Eq. (\ref{parameters}).  For the
isovector channel, we will use the values of the parameters $\alpha_{tv}(\rho_\mathrm{sub})$ and
$\alpha^\prime_{tv}(\rho_\mathrm{sub})$ at the sub-saturation density of 
$\rho_\mathrm{sub}=0.12~\textnormal{fm}^{-3}$.

For the isoscalar-scalar channel we hold on to the  parameter $d_s$, and
express $a_s$, $b_s$ and $c_s$ in terms of the coupling
$\alpha_s(\rho_\mathrm{sat})$ and the corresponding derivatives
\begin{eqnarray}
c_s &=& -\frac{\rho_\mathrm{sat}}{d_s} e^{d_s}\left[ d_s \alpha_s^\prime(\rho_\mathrm{sat})
+ \rho_\mathrm{sat} \alpha_s^{\prime\prime}(\rho_\mathrm{sat}) \right],\\
b_s &=& c_s\left(\frac{1}{d_s}-1 \right) -  \alpha_s^\prime(\rho_\mathrm{sat})
\rho_\mathrm{sat}\frac{e^{d_s}}{d_s},\\
a_s &=& \alpha_s(\rho_\mathrm{sat}) - (b_s+c_s) e^{-d_s} .
\end{eqnarray}
In the isoscalar-vector term there are only three parameters
$a_v$, $b_v$ and $d_v$, which can be expressed as
\begin{eqnarray}
d_v &=& -\frac{\alpha_v^{\prime\prime}(\rho_\mathrm{sat})}{\alpha_v^\prime(\rho_\mathrm{sat})} \rho_\mathrm{sat},\\
b_v &=& - \alpha_v^\prime(\rho_\mathrm{sat})\frac{e^{d_v}}{d_v},\\
a_v &=& \alpha_v(\rho_\mathrm{sat}) - b_v e^{-d_v}.
\end{eqnarray}
Finally, the isovector-vector channel of the functional DD-PC1 is determined by two
parameters only. These are expressed through the values of the coupling
$\alpha_{tv}$ and its derivative $\alpha^\prime_{tv}$ at sub-saturation
density $\rho_\mathrm{sub}=0.12$ fm$^{-3}$:
\begin{eqnarray}
d_{tv} &=& -\rho_\mathrm{sat} \frac{\alpha_{tv}^\prime(\rho_\mathrm{sub})}{\alpha_{tv}(\rho_\mathrm{sub})},\\
b_{tv} &=& \alpha_{tv}(\rho_\mathrm{sub})e^{d_{tv} (\rho_\mathrm{sub}/\rho_\mathrm{sat})}.
\end{eqnarray}

\section{\label{secIV} Infinite and semi-infinite nuclear matter}
The parameters of the functional DD-PC1 given in
Table~\ref{DD-PC1}, or the corresponding terms in the Taylor expansion 
of the density-dependent couplings around the 
saturation point defined in the previous section, correspond to the ``best model'', that 
is, they determine the point $\mathbf{p}_0$ in the $F$-dimensional parameter space. 
In this section we would like to study the ``uniqueness'' of DD-PC1, defined in the sense of 
Ref.~\cite{Fat11}. For this we start with the basic system for which an energy density 
functional can give definite predictions, that is, infinite nuclear matter. In this system 
we define a set of $N$ pseudo-observables ($N > F$) that can be used to compute the 
quality measure $\chi^2(\mathbf{p})$. The model is ``unique'' if all the eigenvalues of 
the  $F \times F$ matrix of second derivatives $\mathcal{M}$ in Eq.~(\ref{M}) are large, 
that is, if all the eigenvectors correspond to stiff directions in the parameter space 
along which the function $\chi^2$ increases rapidly and, therefore, the corresponding linear
combinations of parameters are tightly constrained by the selected observables. 

\begin{table}[htb]
\begin{center}
\caption{\label{Tab:inf-nuc-mat} Pseudo-observables for infinite nuclear matter used to compute the 
quality measure $\chi^2$-function of DD-PC1. The binding energy is evaluated at three density
points:  the saturation density $\rho_0$, at lower density $\rho_{low} = 0.04\;\textnormal{fm}^{-3}$,
and higher density $\rho_{high}=0.56\;\textnormal{fm}^{-3}$. The compressibility modulus $K_0$, 
the Dirac mass $m_D$, the effective mass $m^*$, and the symmetry energy coefficient $a_4$, 
correspond to the saturation density.
The values of the symmetry energy $S_2$ and its slope
 $L$ are given at the sub-saturation density
 $\rho_\mathrm{sub}=0.12\;\textnormal{fm}^{-3}$, which is relevant when calculating binding 
 energies and collective excitations of finite nuclei.
}
\bigskip
\begin{tabular}{c|c} 
\hline
\hline
{\sc observable} & DD-PC1 \\ \hline \hline
$\rho_0$           &  0.152 fm$^{-3}$ \\
$\epsilon(\rho_0)$     & -16.06 MeV \\
$\epsilon(\rho_{low})$  & -6.48 MeV \\
$\epsilon(\rho_{high})$  & 34.38 MeV \\
$K_0$        & 230 MeV \\
$m_D$       & 0.58 \\
$m^*$        & 0.66 \\
 $S_2(\rho_\mathrm{sub})$  & 27.8 MeV \\
 $L(\rho_\mathrm{sub})$  & 57.2 MeV  \\   
 $a_4$ & 33 MeV 
\end{tabular} 
\end{center}

\end{table}

Table \ref{Tab:inf-nuc-mat} lists the values of ten pseudo-observables for infinite nuclear matter 
generated with the functional DD-PC1. In addition to the quantities that are evaluated at the 
saturation point (nucleon density
$\rho_\mathrm{sat}=0.152~\textnormal{fm}^{-3}$, volume energy $a_v=-16.06$
MeV, symmetry energy $a_4 = 33$ MeV, the Dirac mass $m_D = m + \alpha_s \rho_s = 0.58 m$, 
the non-relativistic effective mass \cite{Jam89,Jam90} $m^*=m-\alpha_v \rho_v$ = 0.66 m, 
and the nuclear matter compression modulus $K_{nm} = 230$ MeV), we have included 
four more quantities that characterise the equation of state of symmetric and asymmetric 
matter at lower and higher densities: the binding energy of symmetric nuclear matter at 
low density $\rho_{low} = 0.04\;\textnormal{fm}^{-3}$ 
and high density $\rho_{high}=0.56\;\textnormal{fm}^{-3}$; and both the 
symmetry energy $S_2$ and its slope
$L$ at the sub-saturation density
 $\rho_\mathrm{sub}=0.12\;\textnormal{fm}^{-3}$.
The isoscalar derivative term in Eq.~(\ref{Lagrangian}) does not contribute in the case of 
infinite homogeneous nuclear matter, and thus the surface energy is not included in the 
set of ten pseudo-observables. Since we are considering quantities that cannot be 
directly measured, to calculate the quality function $\chi^2$ and the matrix of second derivatives 
at the point $\mathbf{p}_0$ (DD-PC1) an arbitrary uncertainty of $2\%$ is assigned to
each observable. We are, of course, aware that the empirical values of some of the ten 
pseudo-observables are constrained better than others, e.g. binding energy or compressibility, 
as compared to the slope of symmetry energy. However, this is not crucial for the present 
consideration and thus we prefer to assign the same uncertainty to each observable in 
Table \ref{Tab:inf-nuc-mat}. 
\begin{figure}[htb]
\centering
\begin{tabular}{ccc}
\includegraphics[scale=0.3,trim= 5 35 55 15,clip=true]{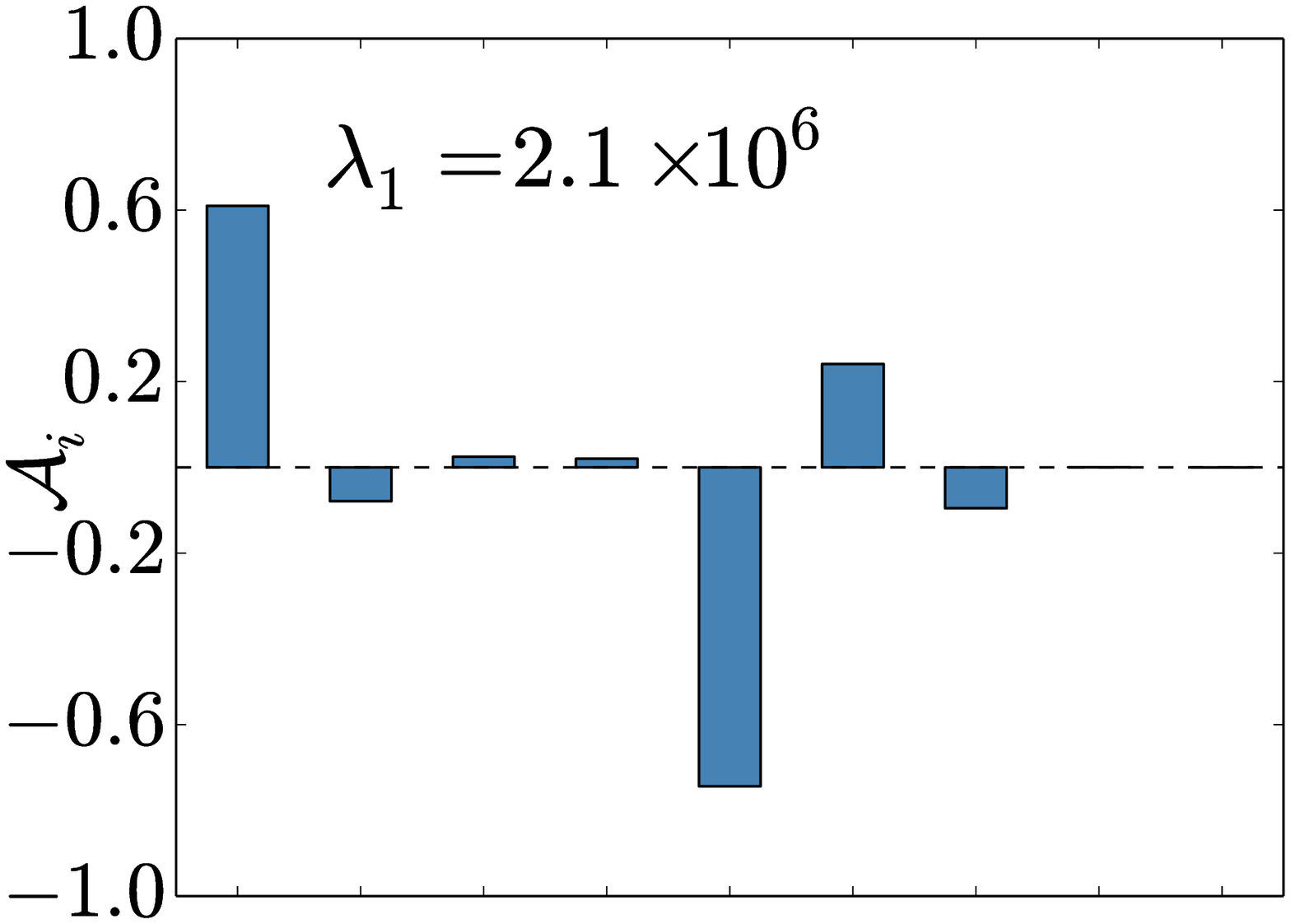} &
\includegraphics[scale=0.3,trim= 60 35 55 15,clip=true]{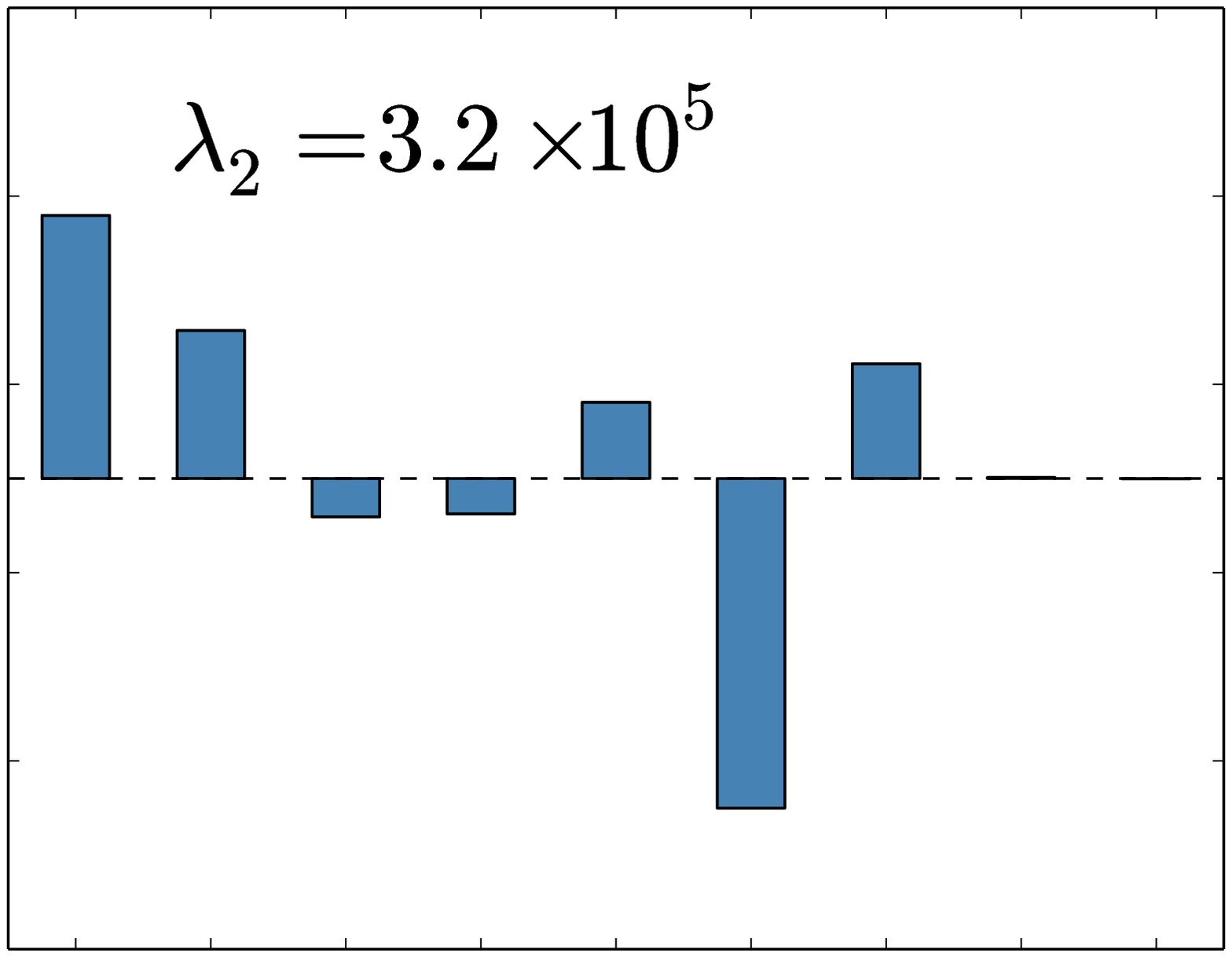} &
\includegraphics[scale=0.3,trim= 60 35 55 15,clip=true]{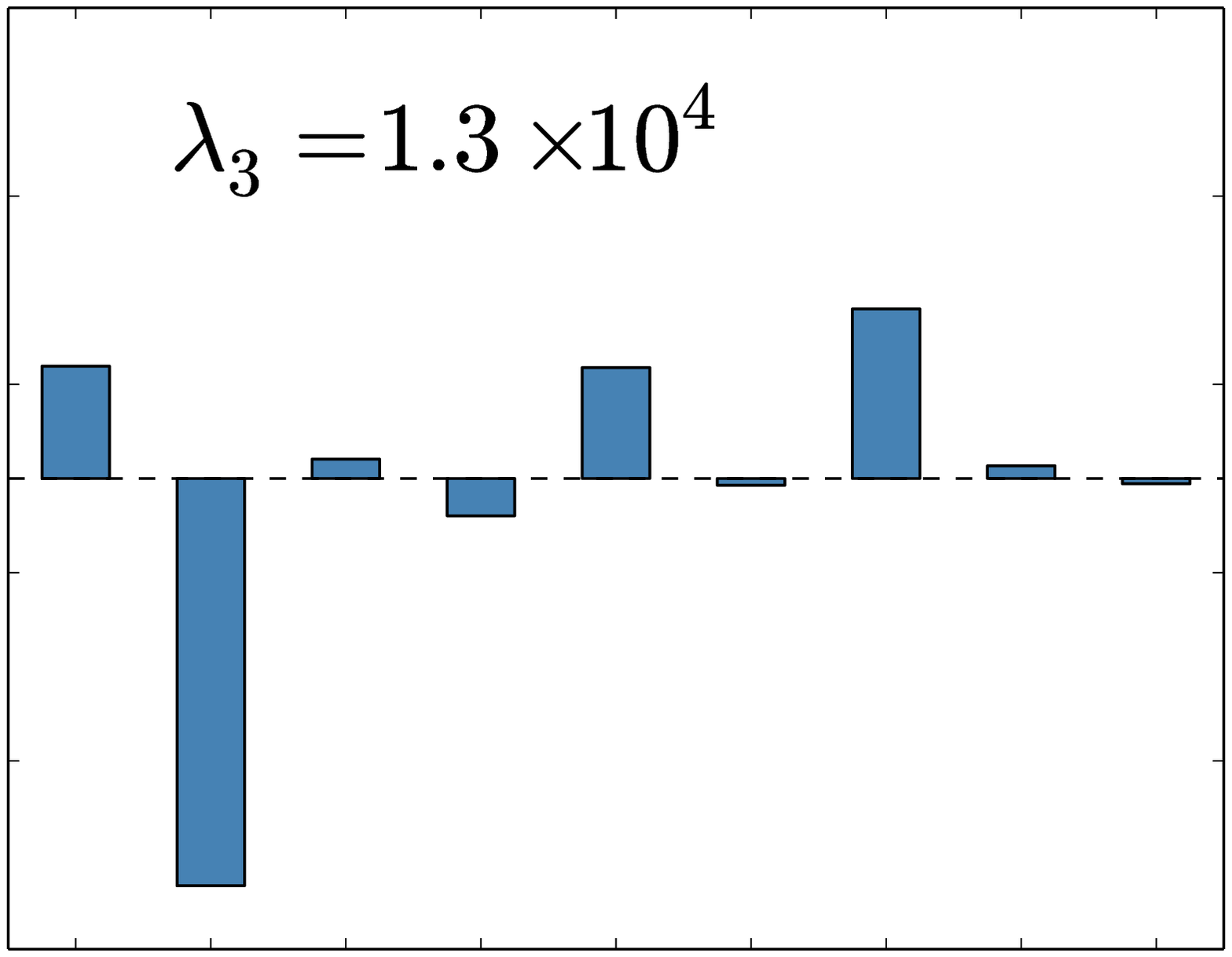} \\
\includegraphics[scale=0.3,trim= 5 35 55 15,clip=true]{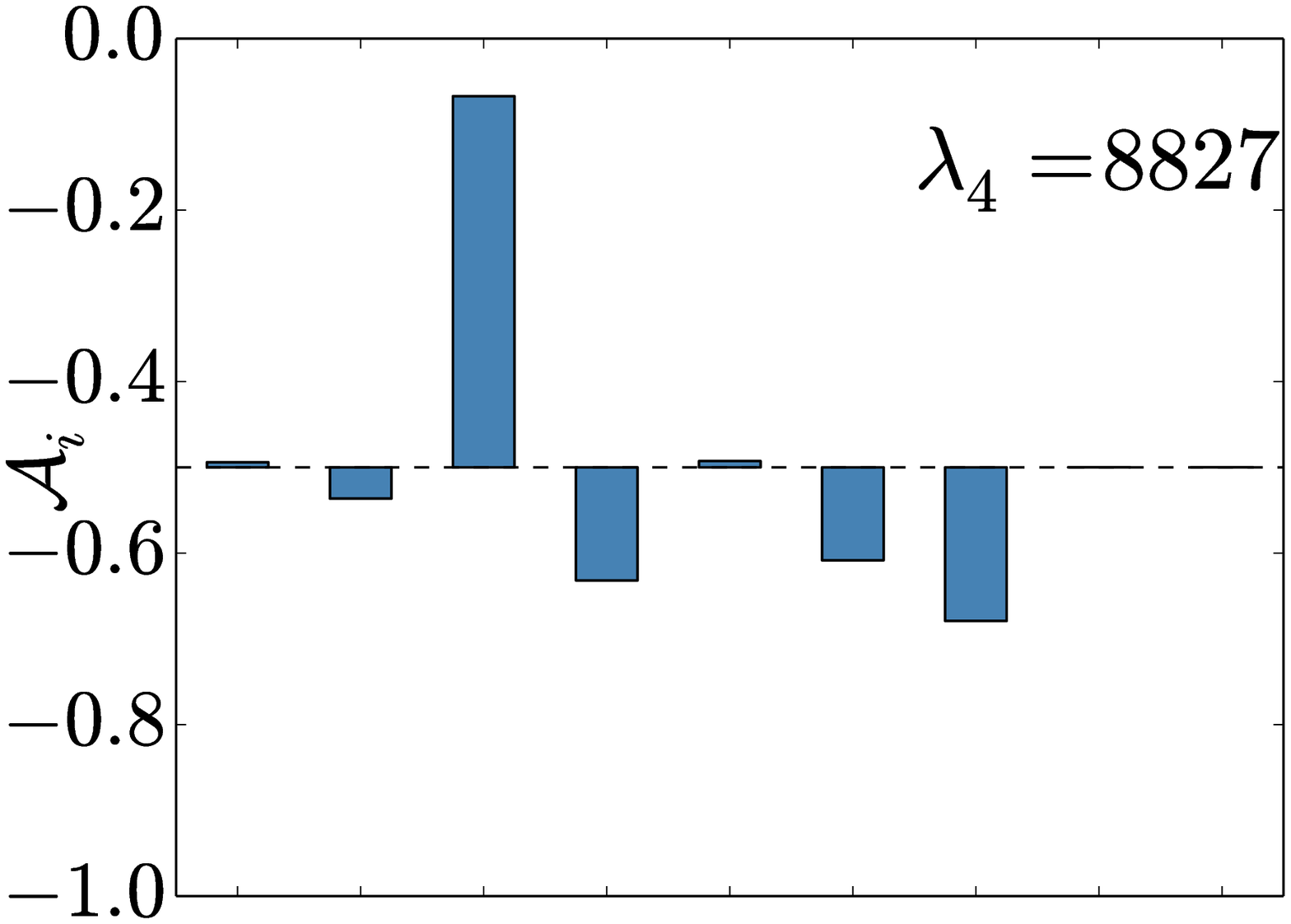} &
\includegraphics[scale=0.3,trim= 60 35 55 15,clip=true]{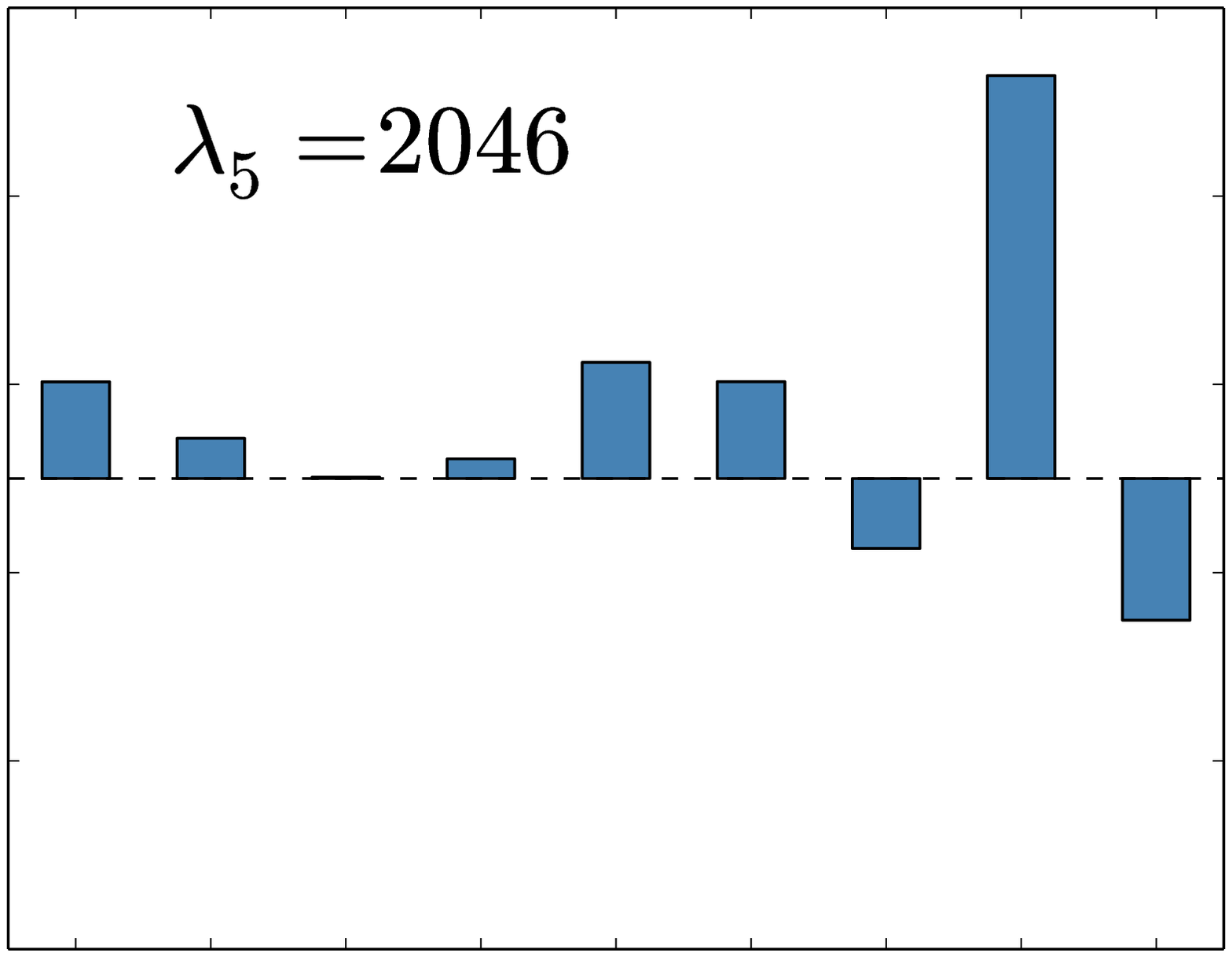} &
\includegraphics[scale=0.3,trim= 60 35 55 15,clip=true]{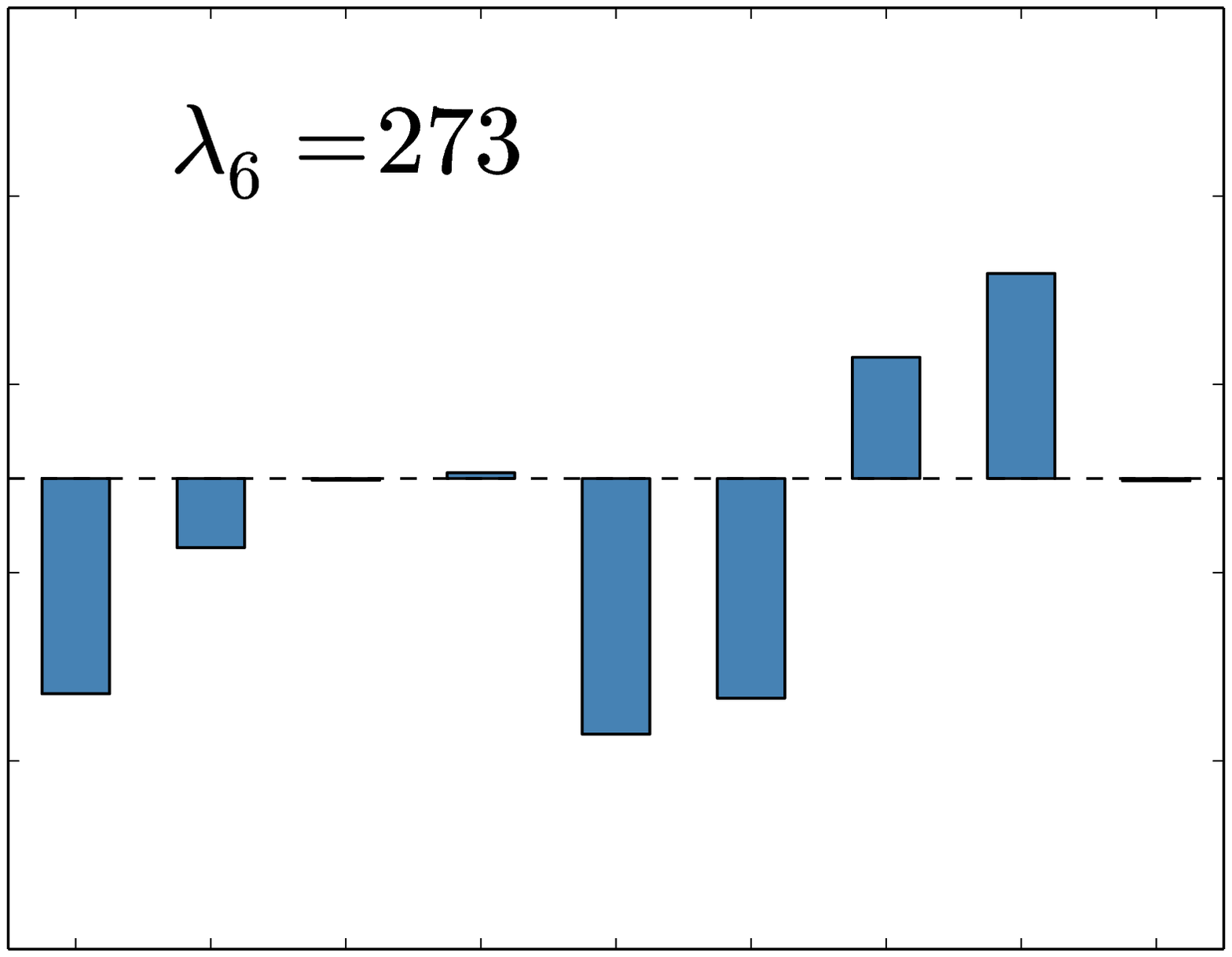} \\
\includegraphics[scale=0.3,trim= 5 0 55 15,clip=true]{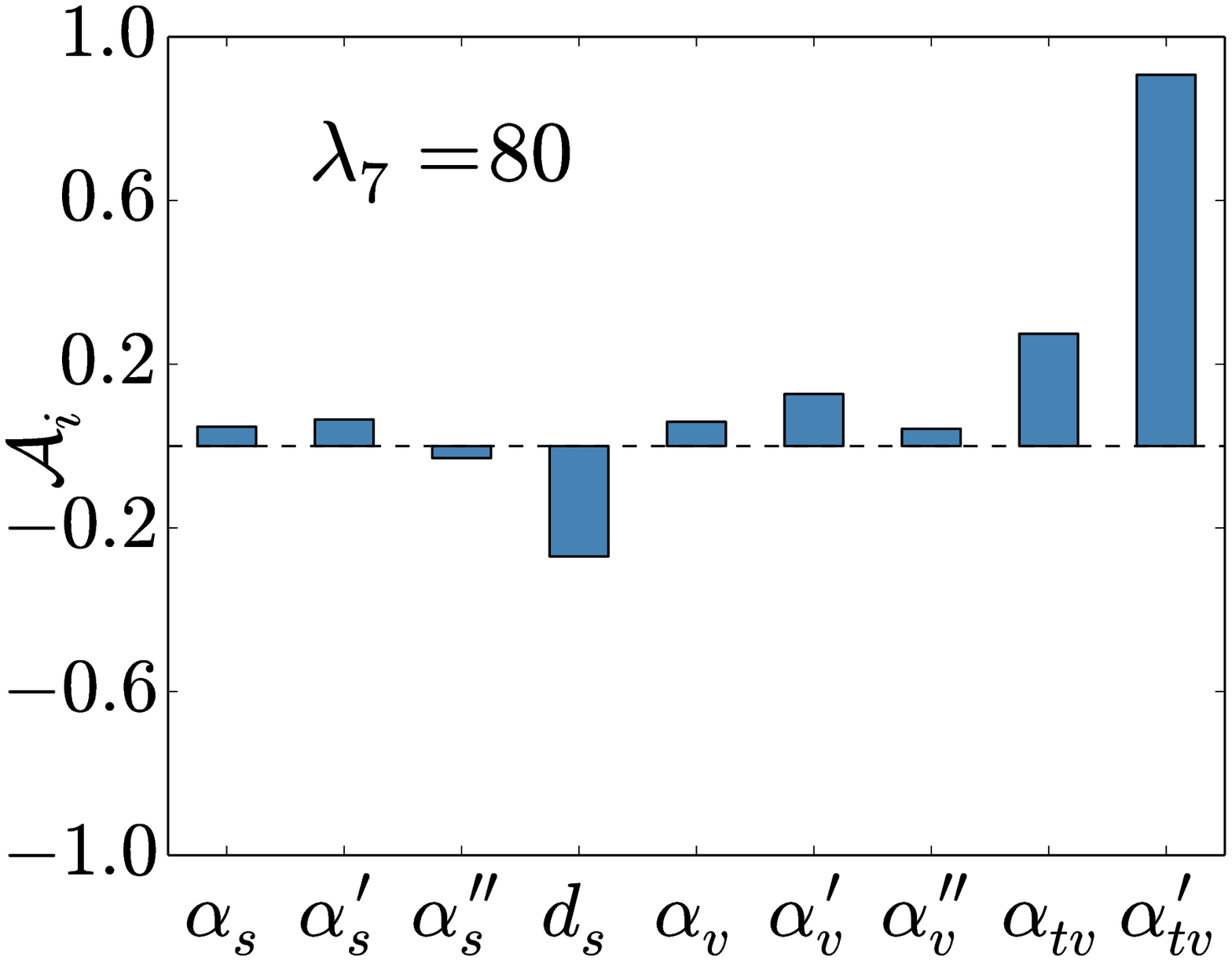} &
\includegraphics[scale=0.3,trim= 60 0 55 15,clip=true]{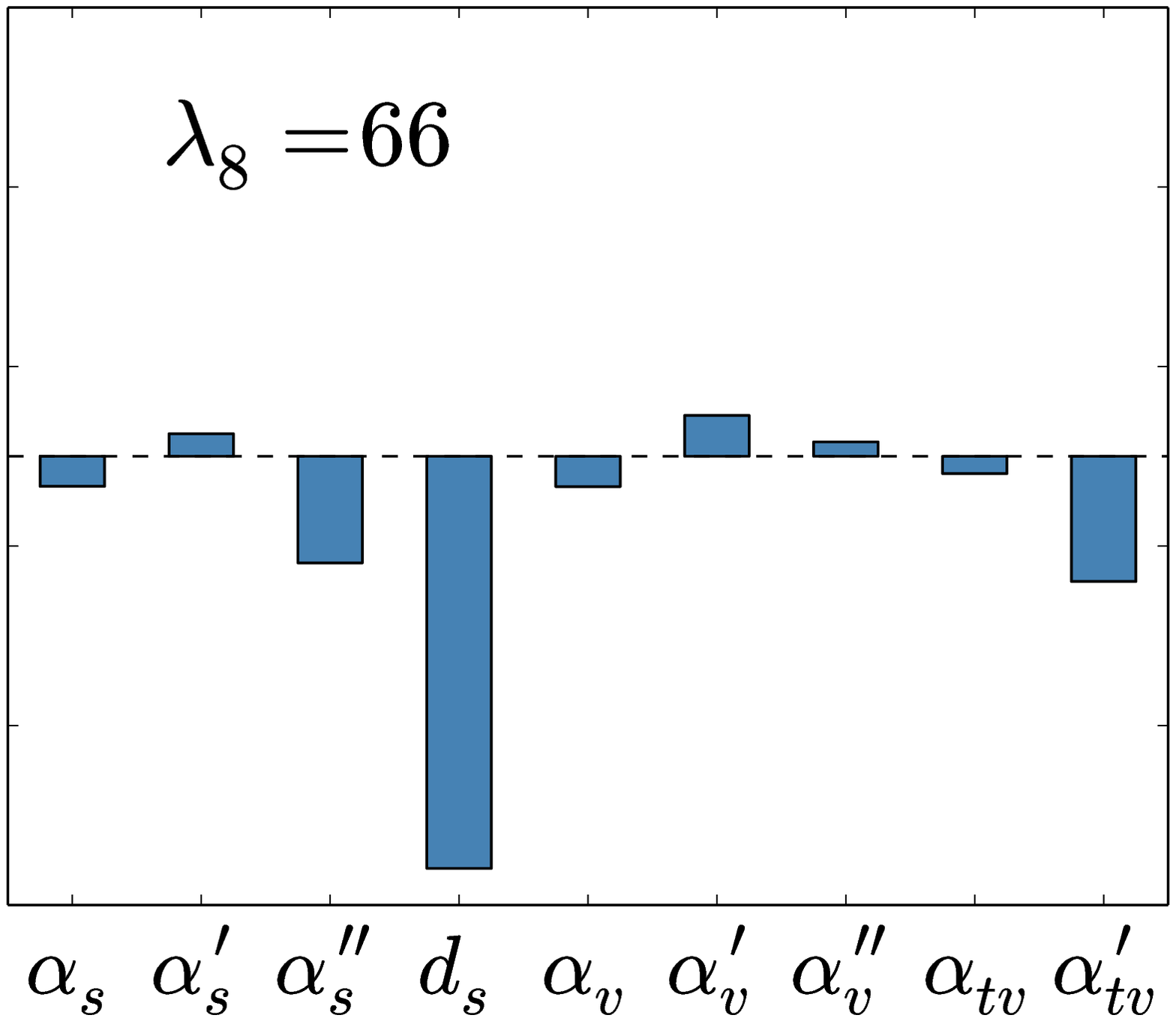} &
\includegraphics[scale=0.3,trim= 60 0 55 15,clip=true]{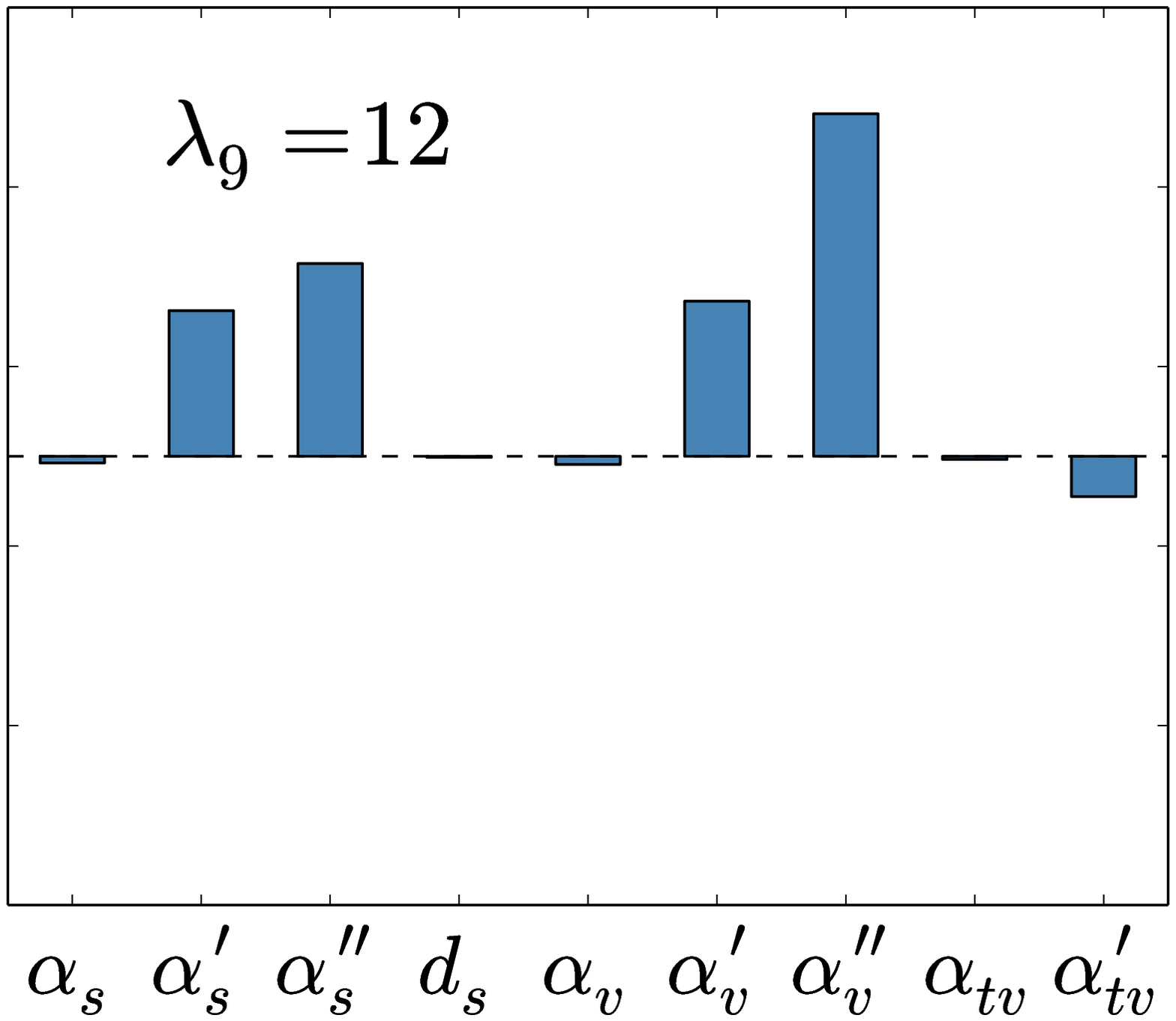}
\end{tabular}
\begin {center}
\caption{\label{fig:mat-A} (Color online) Eigenvalues and eigenvectors of the
$9\times 9$ matrix of second derivatives $\mathcal{M}$ of $\chi^2(\mathbf{p})$ 
in symmetric nuclear matter for the functional DD-PC1.}
\end{center}
\end{figure}

The symmetric $9\times 9$ matrix $\mathcal{M}$ of second derivatives of $\chi^2(\mathbf{p})$ 
at the point $\mathbf{p}_0$ (DD-PC1) is diagonalized by means of an orthogonal transformation. The diagonal 
matrix elements in order of decreasing values 
and the components of the corresponding eigenvectors are displayed in Fig.~\ref{fig:mat-A}.
Stiff directions in the nine-parameter space are characterised by large eigenvalues, that is,
the function $\chi^2$ increases rapidly along these directions. This means that the particular 
linear combinations of parameters corresponding to the stiff eigenvectors are firmly determined 
by the pseudo-data listed in Table \ref{Tab:inf-nuc-mat}.
On the other hand, comparatively small eigenvalues belong to soft directions in the multi-parameter space, 
along which the quality measure displays little deterioration and the corresponding 
linear combinations of parameters that define the energy density functional are poorly 
constrained. 

The four stiffest directions in Fig.~\ref{fig:mat-A} are dominated by isoscalar parameters, 
as denoted by the components of the corresponding eigenvectors,
whereas the fifth, sixth and seventh eigenvector contain sizeable admixtures of isovector parameters. 
The two softest directions are again predominantly isoscalar. The first mode, characterised by the 
largest eigenvalue, corresponds to out-of-phase oscillations of the $\alpha_s(\rho_\mathrm{sat})$ and
$\alpha_v(\rho_\mathrm{sat})$ coupling parameters. This combination of parameters is tightly 
constrained by the three values of the nuclear matter binding
energy. An increase of the scalar attraction and a simultaneous decrease of the vector repulsion 
leads to a pronounced increase of the binding energy, as described in the previous section and, 
therefore, to a rapid deterioration of $\chi^2$. Mode two, which corresponds to out-of-phase oscillations of the 
derivatives $\alpha_s^\prime(\rho_\mathrm{sat})$ and $\alpha_v^\prime(\rho_\mathrm{sat})$, is 
predominantly constrained by the values of the binding energy at $\rho_{low}$ and $\rho_{high}$
because the slope of the couplings at saturation density determines the values
of the corresponding couplings below and above the saturation density. 
Mode three is mostly determined by the saturation density, whereas for mode four 
the largest amplitudes correspond to components that represent the second derivatives of 
isoscalar couplings, that is, this mode is constrained by the nuclear matter 
incompressibility.

Modes number five and six correspond to superpositions of the isoscalar and isovector modes, and are 
constrained by the value of $S_2$  and the Dirac mass, which enters into the expression for 
the symmetry energy
\begin{equation}
S_2(\rho) = \frac{k_f^2}{6\sqrt{k_f^2+{m_D}^2}} + \frac{1}{2} \alpha_{tv}(\rho) \rho\; .
\end{equation}
Because we consider the value of the symmetry energy at sub-saturation density, both the couplings and
their derivatives ($\alpha_s(\rho_\mathrm{sat})$, $\alpha_s^\prime(\rho_\mathrm{sat})$,
$\alpha_v(\rho_\mathrm{sat})$, $\alpha_v^\prime(\rho_\mathrm{sat})$) contribute to these modes.
When the slope of the symmetry energy is calculated, both the coupling $\alpha_{tv}(\rho)$
and its derivative $\alpha_{tv}(\rho)$ contribute, as signaled by the in-phase
isovector components in mode seven.

Mode number eight corresponds almost entirely to the parameter $d_s$ in the isoscalar-scalar 
channel (cf. Eq.~(\ref{parameters})), which is obviously poorly
determined by the pseudo-data in Tab.~\ref{Tab:inf-nuc-mat}. The softest mode represents
the in-phase contributions from the first and second derivatives of the isoscalar couplings 
at saturation density.

\begin{figure}[htb]
\centering

\includegraphics[scale=0.4,trim= 0 0 50 15,clip=true]{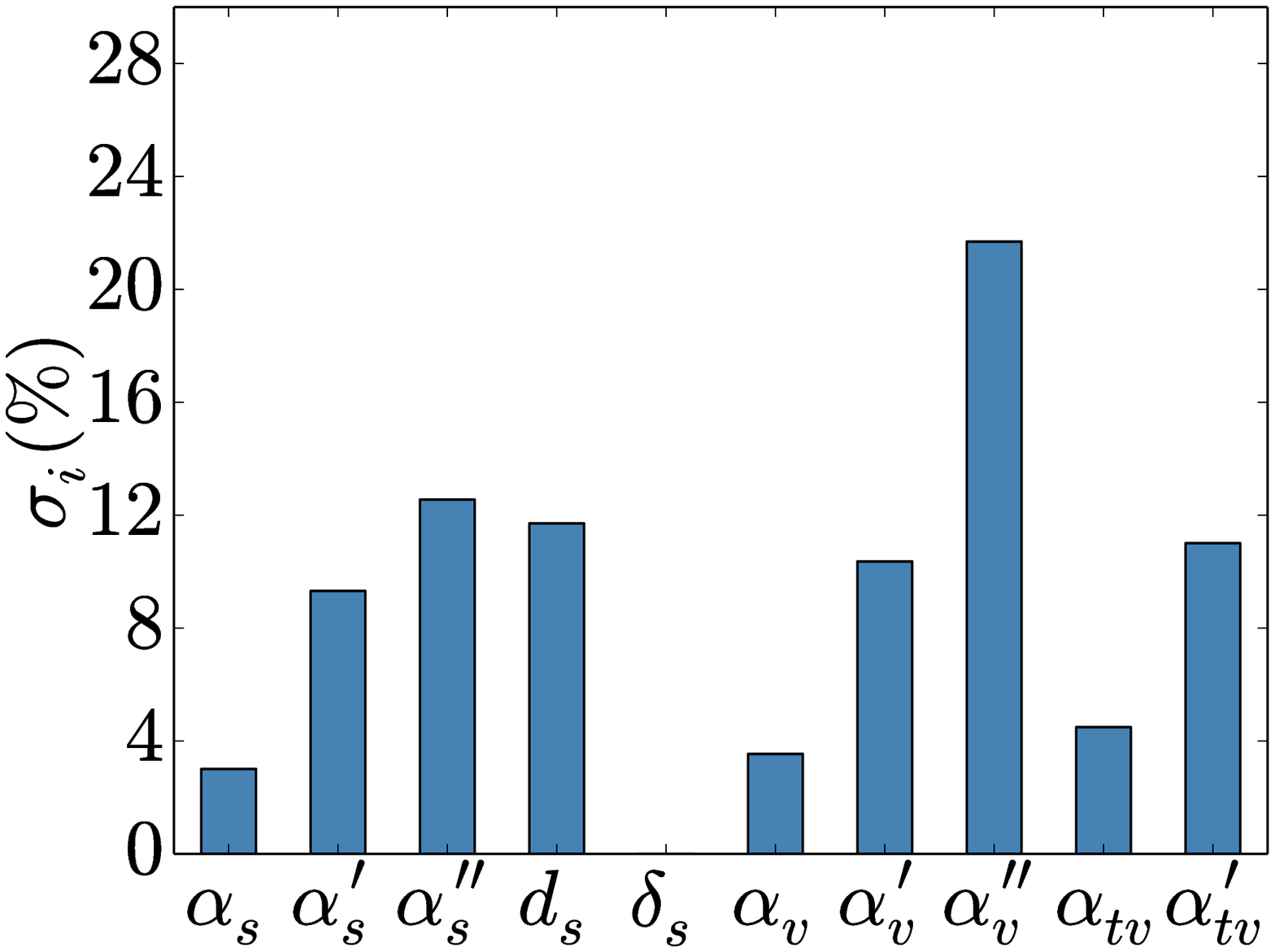} 
\includegraphics[scale=0.4,trim= 0 0 50 15,clip=true]{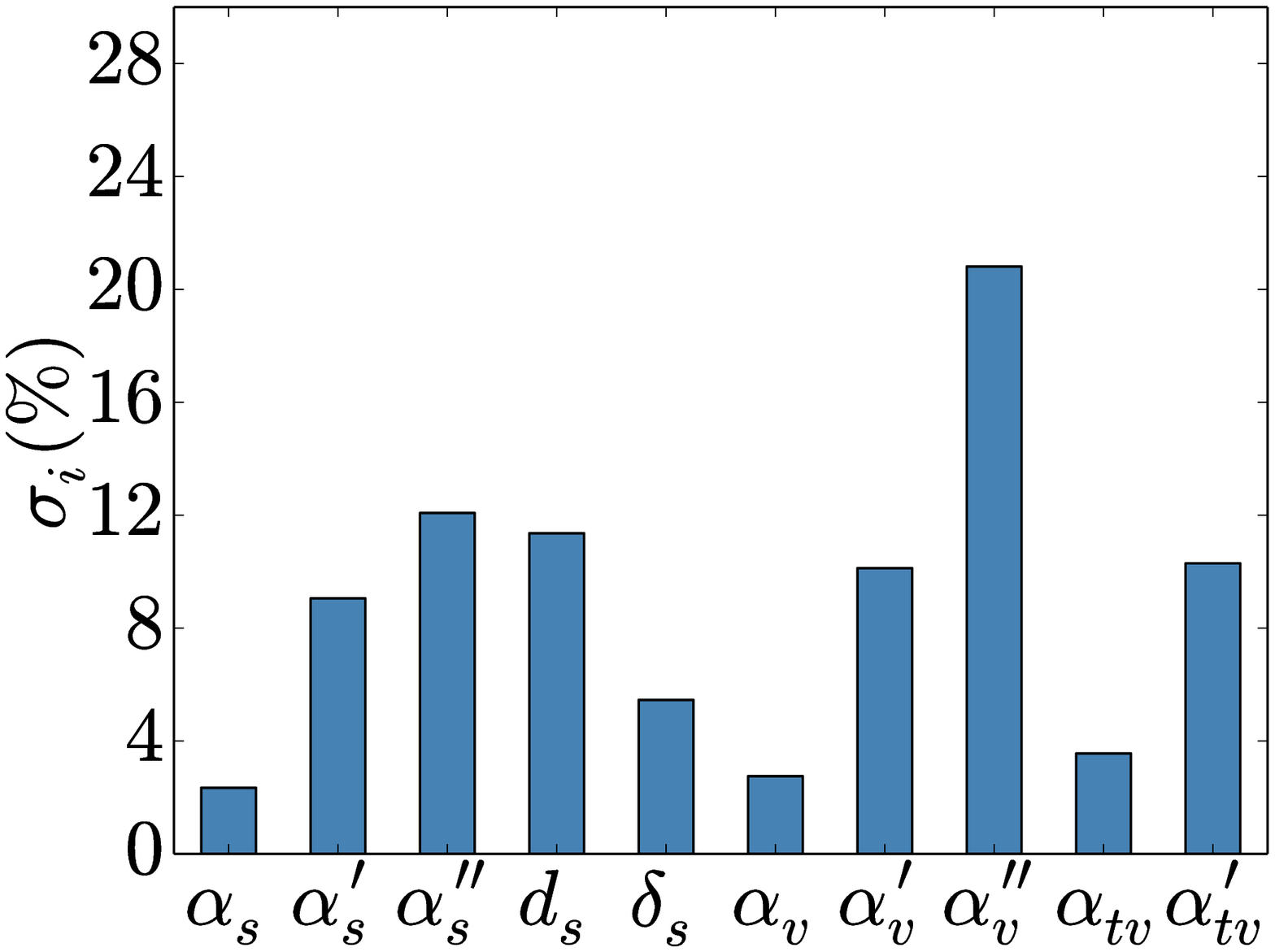} 
\begin{center}
\caption{\label{fig:uncertainties-parameters} (Color online) Uncertainties $\sigma_i$ in percentage 
of the model parameters for the functional DD-PC1. Results displayed in the left panel correspond to a 
calculation with pseudo-observables of infinite nuclear matter, 
while those in the right panel have been obtained by including also the surface energy of semi-infinite
nuclear matter. }
\end{center}
\end{figure}

The uncertainties, that is, the variances of model parameters are given by the 
diagonal elements of the inverse matrix $\mathcal{M}^{-1}$ of second derivatives of $\chi^2$ -- 
the covariance matrix 
(cf. Eq.~(\ref{eq:mixvariances}))
\begin{equation}
\sigma_i^2 = \left( \mathcal{M}^{-1} \right)_{ii} 
= \left(\mathcal{A} \mathcal{D}^{-1}\mathcal{A}^T  \right)_{ii} 
= \sum_{j=1}^9{\mathcal{A}_{ij}\lambda_j^{-1}} .
\end{equation}
For each parameter $p_i = p_{0i} (1\pm \sigma_i)$, the uncertainty $\sigma_i$ in percentage is 
shown in the left panel of Fig.~\ref{fig:uncertainties-parameters}. As one would have expected,
the values of the couplings $\alpha_s(\rho_\mathrm{sat})$, $\alpha_v(\rho_\mathrm{sat})$ and
$\alpha_{tv}(\rho_\mathrm{sub})$ have the smallest uncertainties 
($\leq 5\%$), whereas uncertainties increase rapidly for their first and second   
derivatives. This result already indicates that a particular choice for the density dependence of 
coupling functions can lead to large model uncertainties. As shown in Fig.~\ref{fig:mat-A}, 
the parameter $d_s$ completely determines the very soft mode number eight, and the corresponding 
uncertainty of this parameter is rather large ($> 10\%$). The uncertainty of $\delta_s$ is zero 
because the corresponding term does not contribute to homogeneous nuclear matter.  

\begin{figure}[htb]
\centering
\begin{tabular}{cc}
\includegraphics[scale=0.4,trim= 25 0 0 15,clip=true]{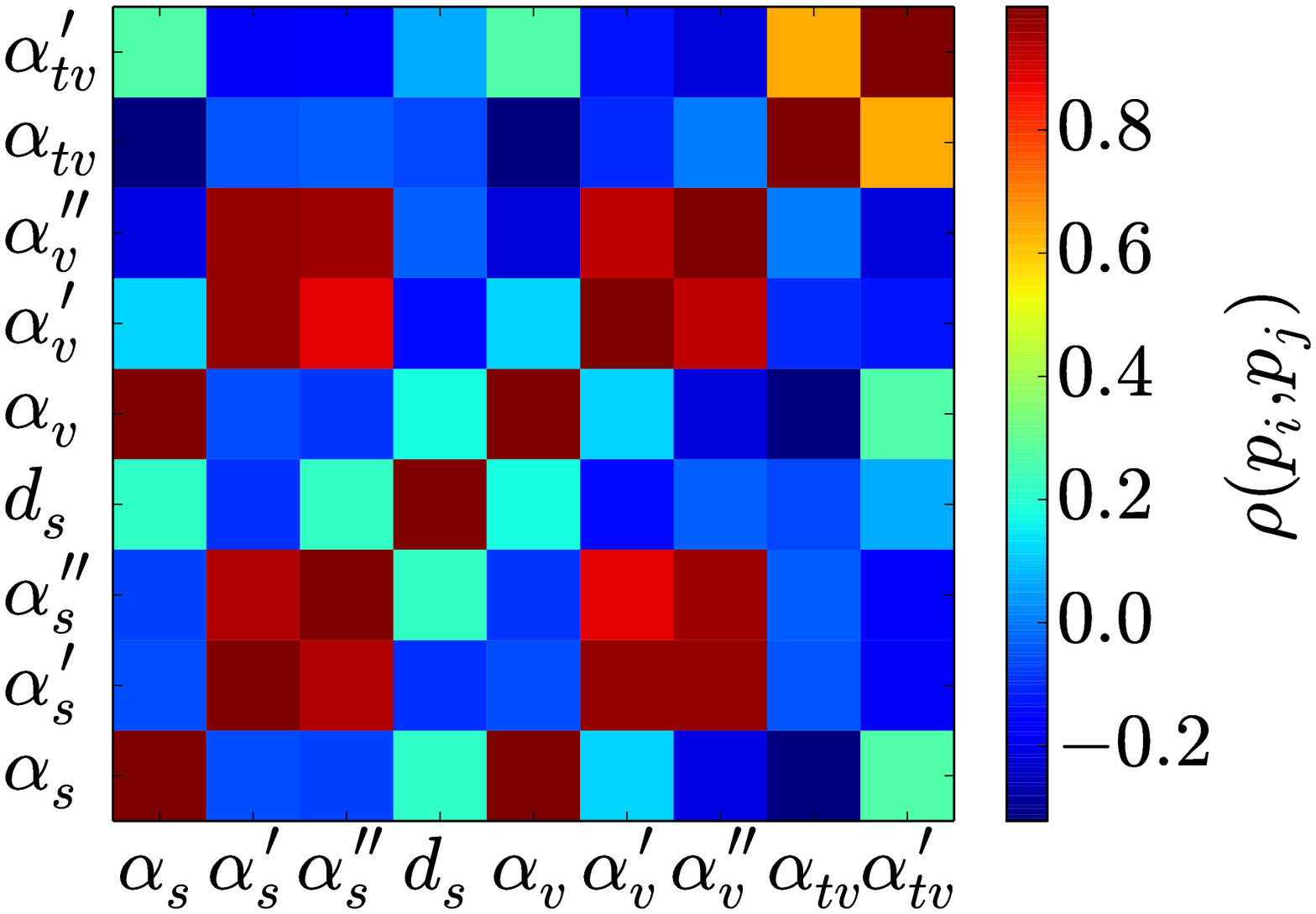} &
\includegraphics[scale=0.4,trim= 25 0 0 15,clip=true]{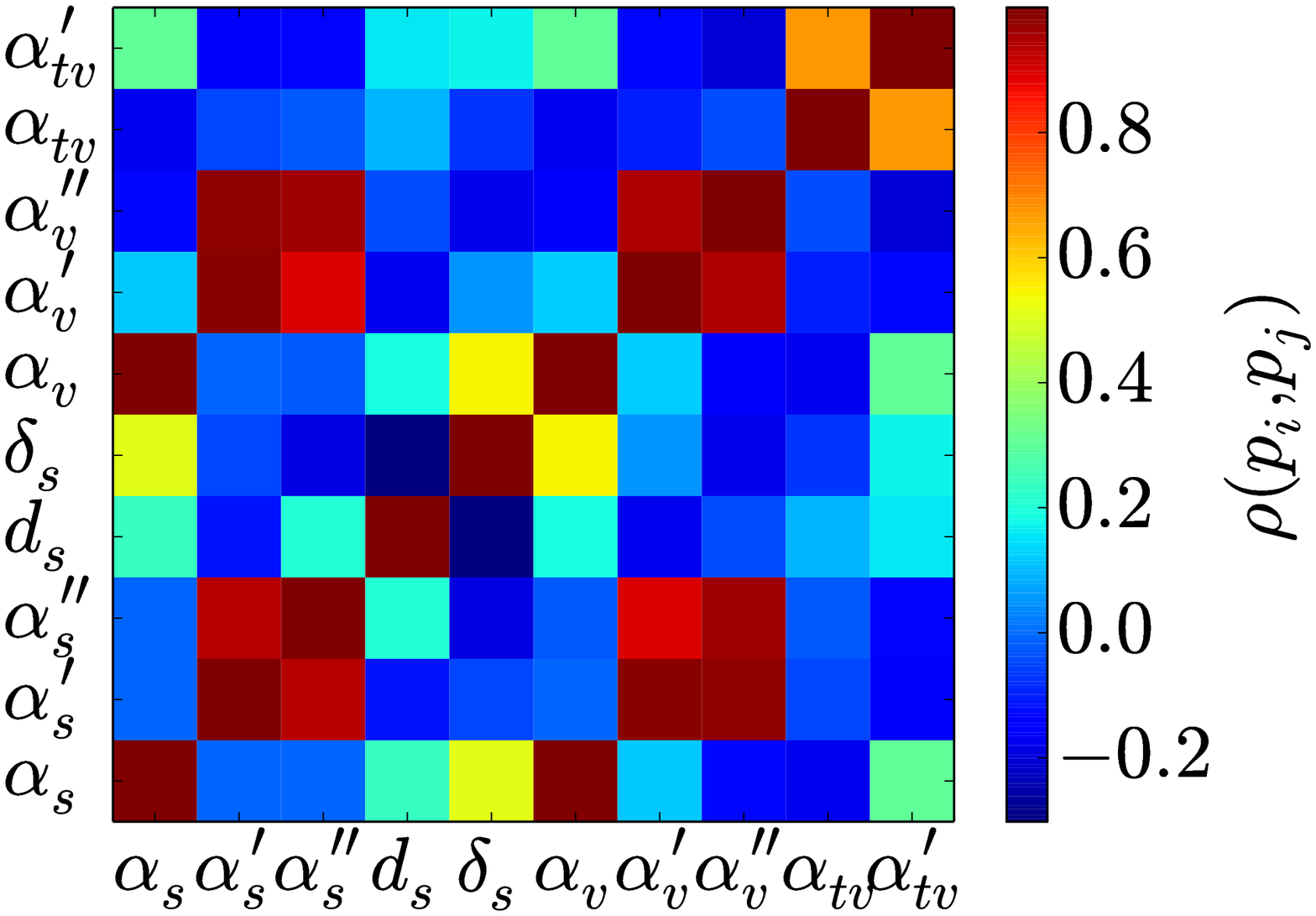} 
\end{tabular}
\caption{\label{fig:correlations-parameters} (Color online) Left: color-coded plot of the 36
independent correlation coefficients between the 9 model parameters that contribute 
to the calculation for infinite nuclear matter. Right: 
color-coded plot of the 45
independent correlation coefficients between the 10 model parameters that 
contribute when semi-infinite
nuclear matter is included in the calculation.}
\end{figure}
The correlation coefficients between model parameters are determined by the off-diagonal 
elements of the covariance matrix:
\begin{equation}
\rho(p_i,p_j) = \frac{\mathcal{M}^{-1}_{ij}}{\sqrt{ {\mathcal{M}}^{-1}_{ii} {\mathcal{M}}^{-1}_{jj}}}\; .
\end{equation}
The color coded plot of the 36 independent correlation
coefficients for the present calculation of infinite nuclear matter is displayed in the left panel of 
Fig.~\ref{fig:correlations-parameters}. 
One notices the strong correlations between the isoscalar scalar and isoscalar vector 
couplings, as well as between their first derivatives and also second derivatives. 
There is also a significant
correlation between $\alpha_{tv}(\rho_\mathrm{sub})$ and $\alpha^\prime_{tv}(\rho_\mathrm{sub})$
because the coupling $\alpha_{tv}(\rho_\mathrm{sub})$ enters
the expression for the slope of the symmetry energy. The correlation between
the isoscalar and isovector parameters originates from the Dirac mass contribution
to the symmetry energy.

\begin{figure}[htb]
\centering
\begin{tabular}{cc}
\includegraphics[scale=0.4,trim= 0 0 55 15,clip=true]{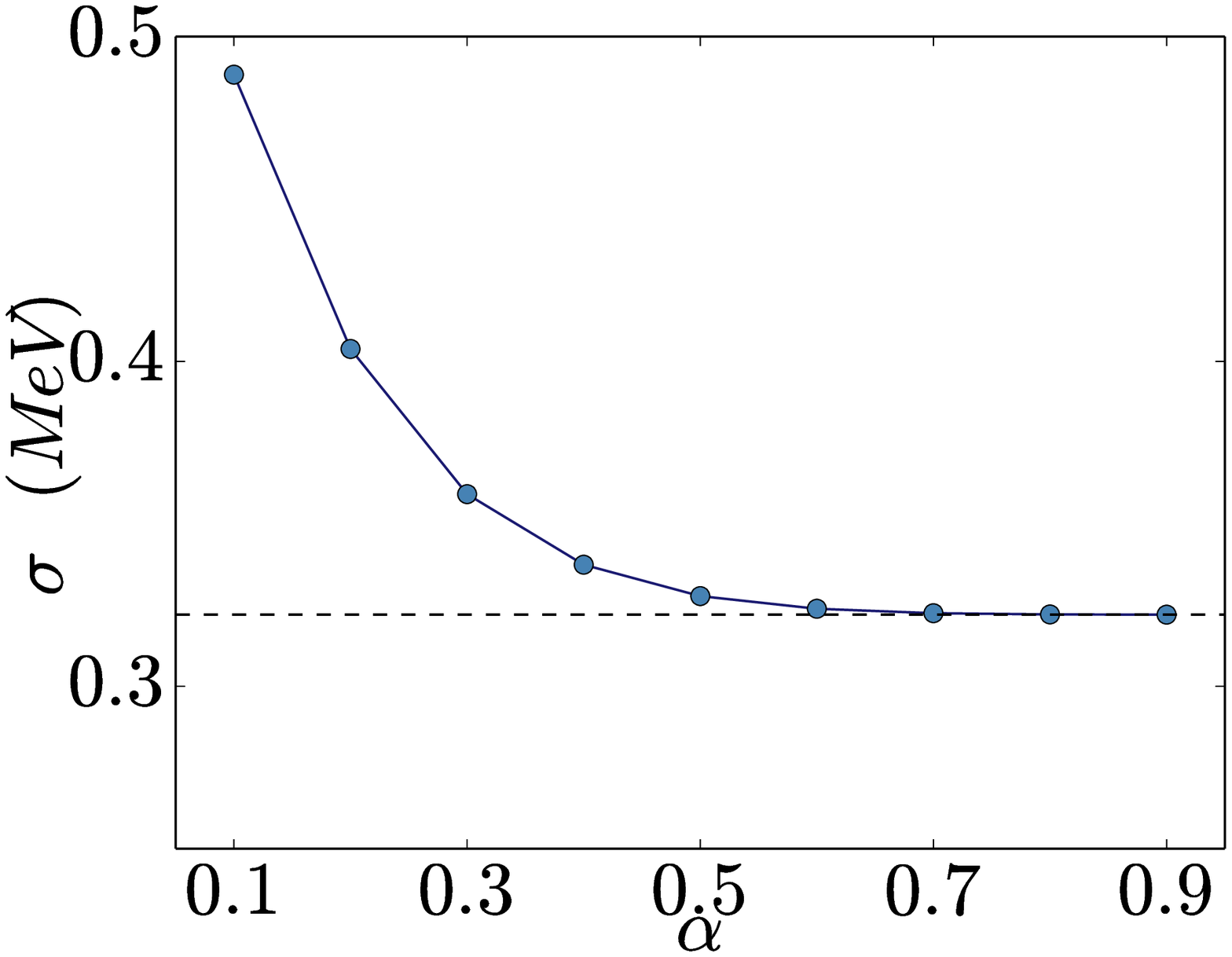} &
\includegraphics[scale=0.4,trim= 0 0 55 15,clip=true]{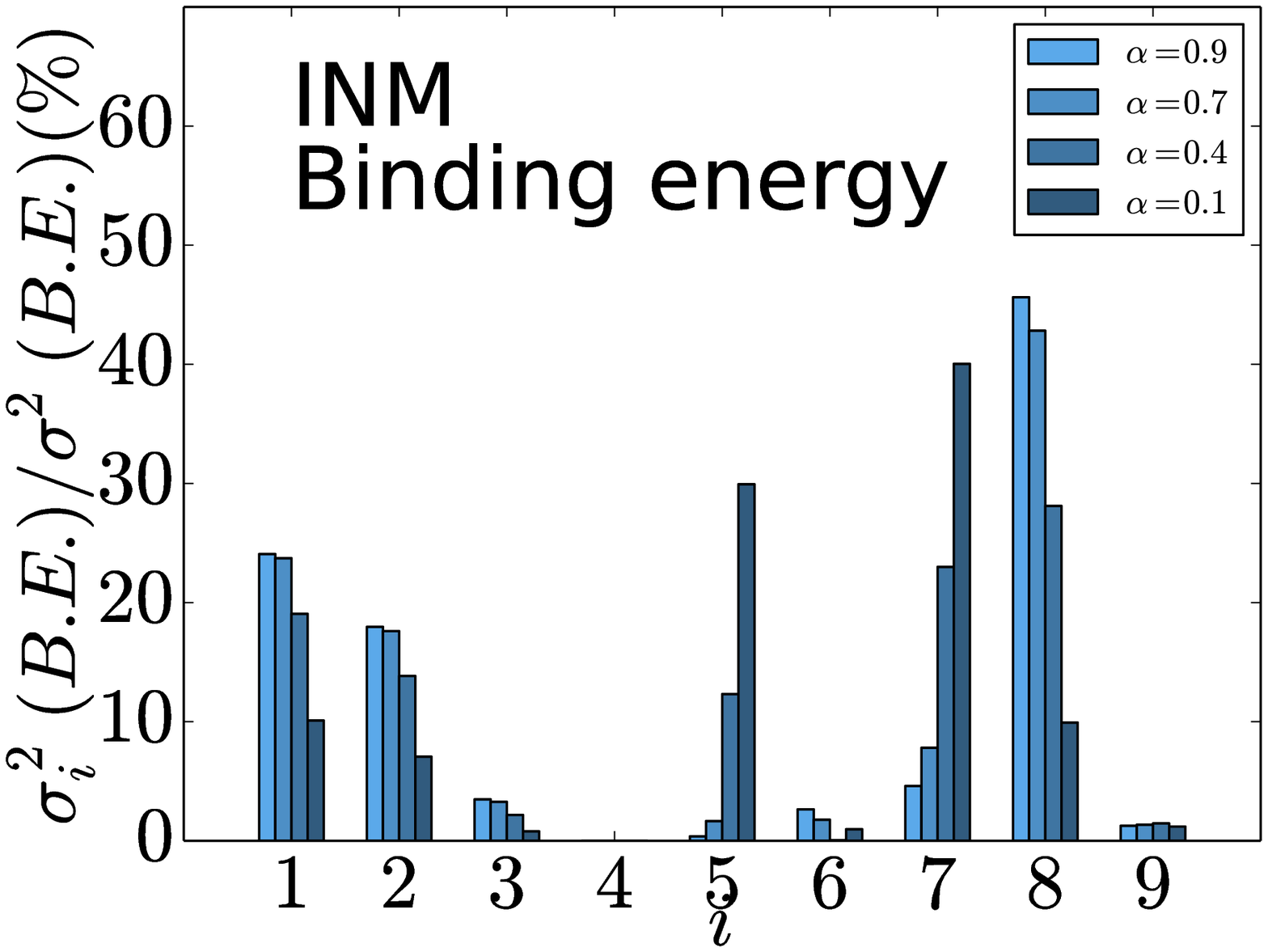}
\end{tabular} 
\begin{center}
\caption{\label{fig:sigma-asymmetry} (Color online) Left: calculated uncertainty of the binding
energy of asymmetric nuclear matter at the saturation density $\rho_0=0.152$ fm$^{-3}$, 
as a function of the asymmetry parameter. Right: relative contributions in percentage 
of the nine linear combinations of model parameters that correspond to the eigenvectors of the 
matrix of second derivatives $\mathcal{M}$ in Eq.~(\ref{M}) to the variance of the binding energy of 
asymmetric nuclear matter.}
\end{center}
\end{figure}

In the left panel of Fig.~\ref{fig:sigma-asymmetry} we plot the
calculated uncertainty $\sigma(\epsilon(\rho_0)) = \sqrt{\textnormal{var}(\epsilon(\rho_0))}$ of the binding energy 
of asymmetric nuclear matter at the saturation density $\rho_0=0.152$ fm$^{-3}$, as a function of the 
asymmetry parameter. The asymmetry parameter $\alpha$ is defined
by the relation
\begin{equation}
\alpha = \frac{\rho_p}{\rho_n} \Longrightarrow
\rho_n = \frac{\rho}{1+\alpha}, \quad \rho_p = \rho-\rho_n = \frac{\alpha \rho}{1+\alpha}\;.
\end{equation}
For symmetric nuclear matter the asymmetry parameter equals one, and it goes to zero for 
pure neutron matter. The dashed horizontal line corresponds to the assumed 2\% uncertainty in 
symmetric nuclear matter. The calculated uncertainty exhibits a rapid increase when nuclear 
matter becomes neutron rich and, in finite nuclei, this would result in a corresponding 
increase of the standard error of calculated binding energies for systems with a large 
neutron excess, as shown in the analysis of the propagation of uncertainties in Skyrme 
energy density functionals \cite{Gao13}. For four values of $\alpha$, 
the right panel of Fig.~\ref{fig:sigma-asymmetry} displays 
the individual relative contributions from each of the nine linear combinations of 
model parameters, that correspond to the eigenvectors of the 
matrix of second derivatives $\mathcal{M}$ in Eq.~(\ref{M}), to the variance of the binding energy of 
asymmetric nuclear matter. Here one can clearly identify the source of the increase in uncertainty 
$\sigma(\epsilon(\rho_0))$ with decreasing $\alpha$. As the neutron excess increases, 
we notice a pronounced increase in the relative contributions of the eigenvectors (modes) 
five and seven (see Fig.~\ref{fig:mat-A}) to the variance of the binding energy. These two 
modes are dominated by the two isovector parameters: the values of the coupling
$\alpha_{tv}$ and its derivative $\alpha^\prime_{tv}$ at sub-saturation
density, respectively. The uncertainties of the isovector parameters, especially of the 
slope of the symmetry energy, determine the pronounced increase of 
the uncertainty of the binding energy in asymmetric nuclear matter.

\begin{figure}[htb]
\centering
\begin{tabular}{ccc}
\includegraphics[scale=0.3,trim= 5 35 55 15,clip=true]{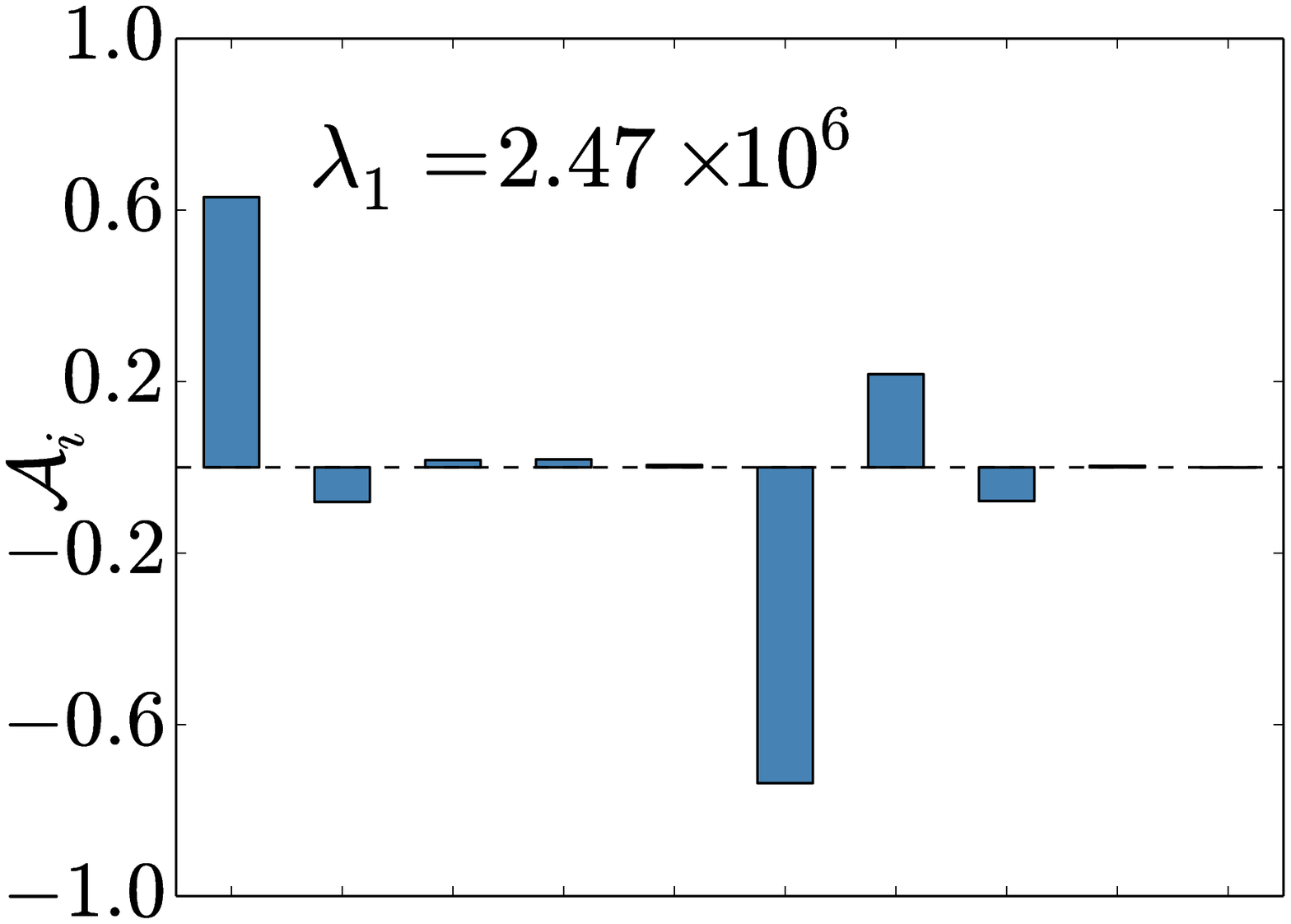}& & \\
\includegraphics[scale=0.3,trim= 5 35 55 15,clip=true]{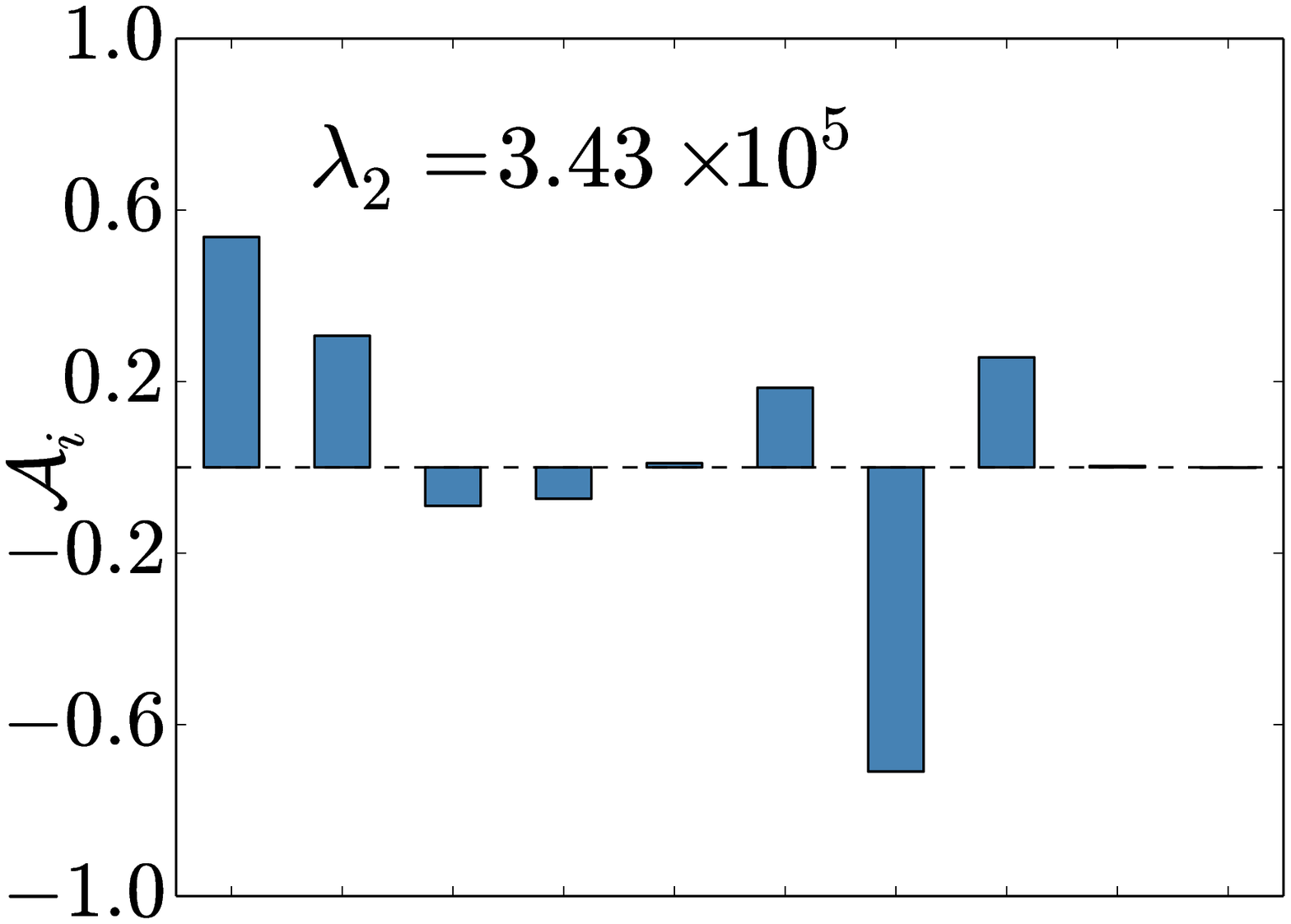} &
\includegraphics[scale=0.3,trim= 60 35 55 15,clip=true]{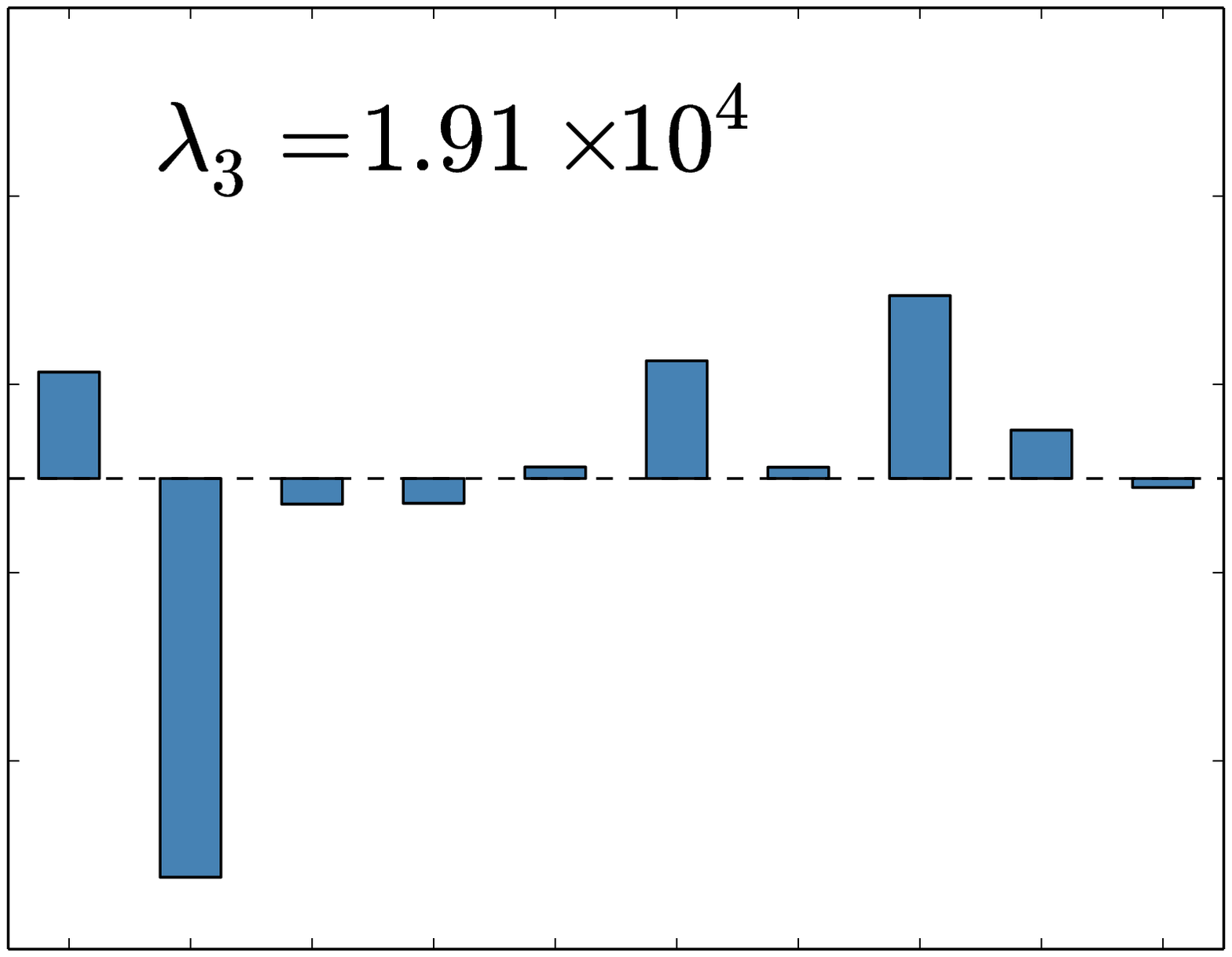} &
\includegraphics[scale=0.3,trim= 60 35 55 15,clip=true]{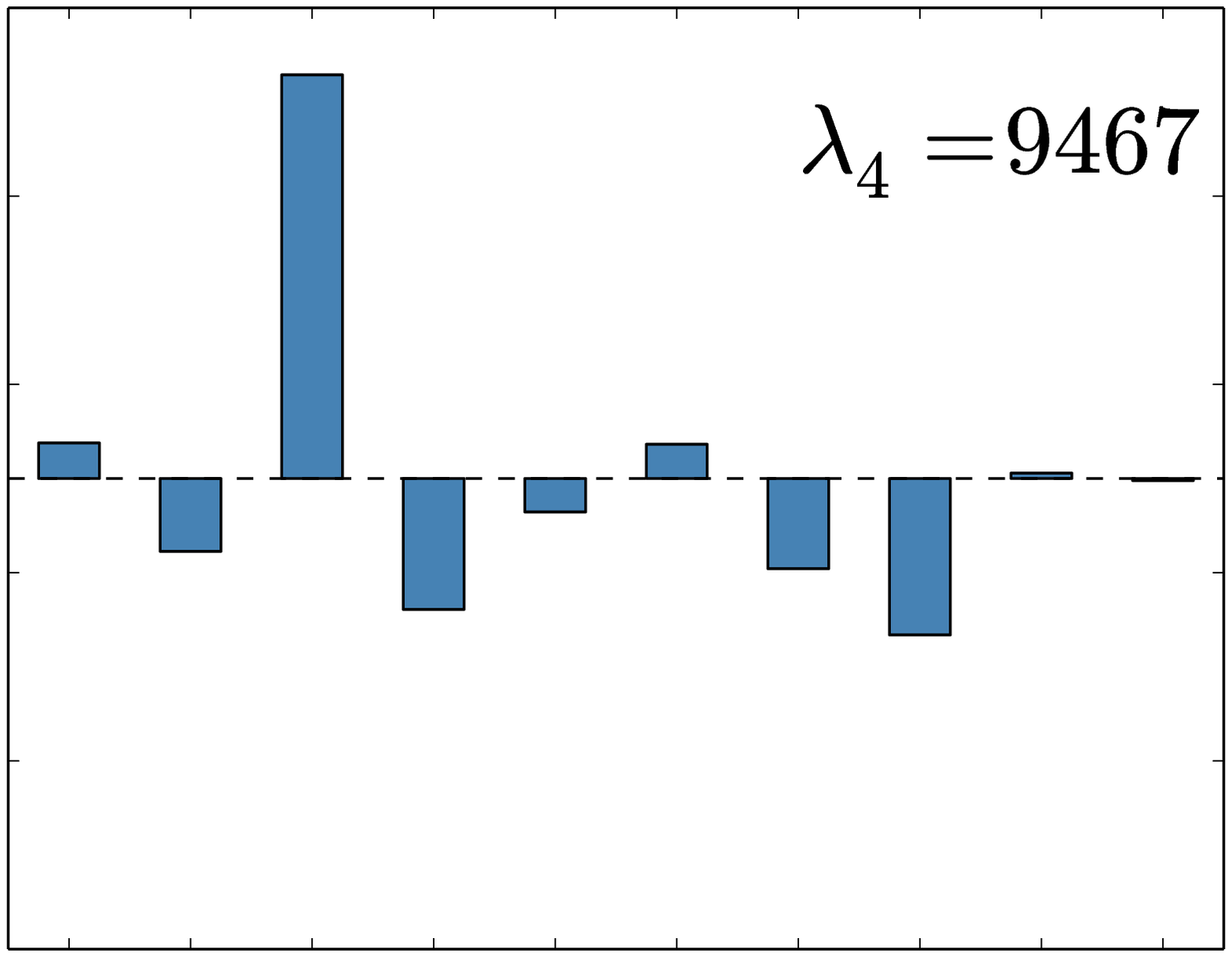} \\
\includegraphics[scale=0.3,trim= 5 35 55 15,clip=true]{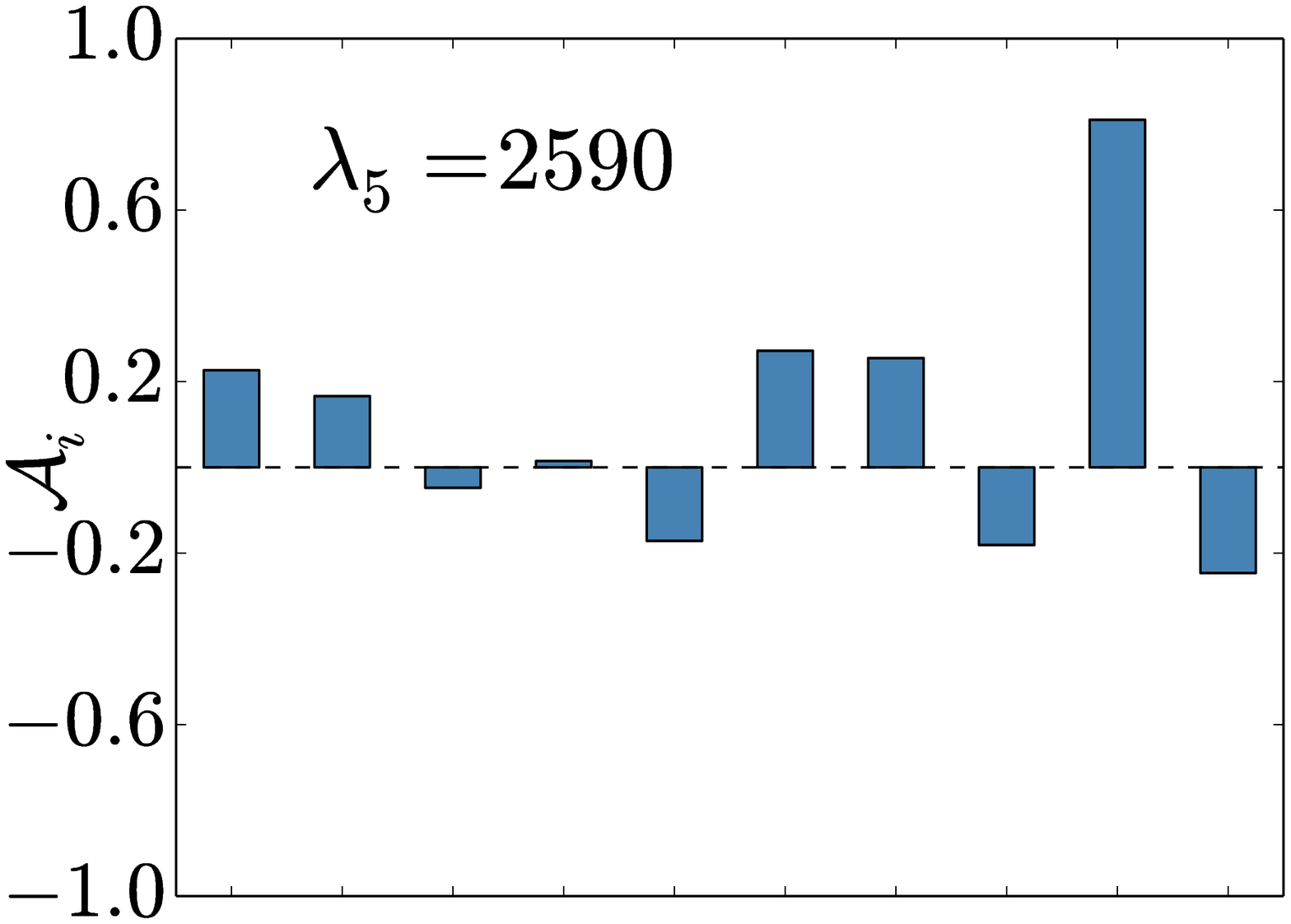}&
\includegraphics[scale=0.3,trim= 60 35 55 15,clip=true]{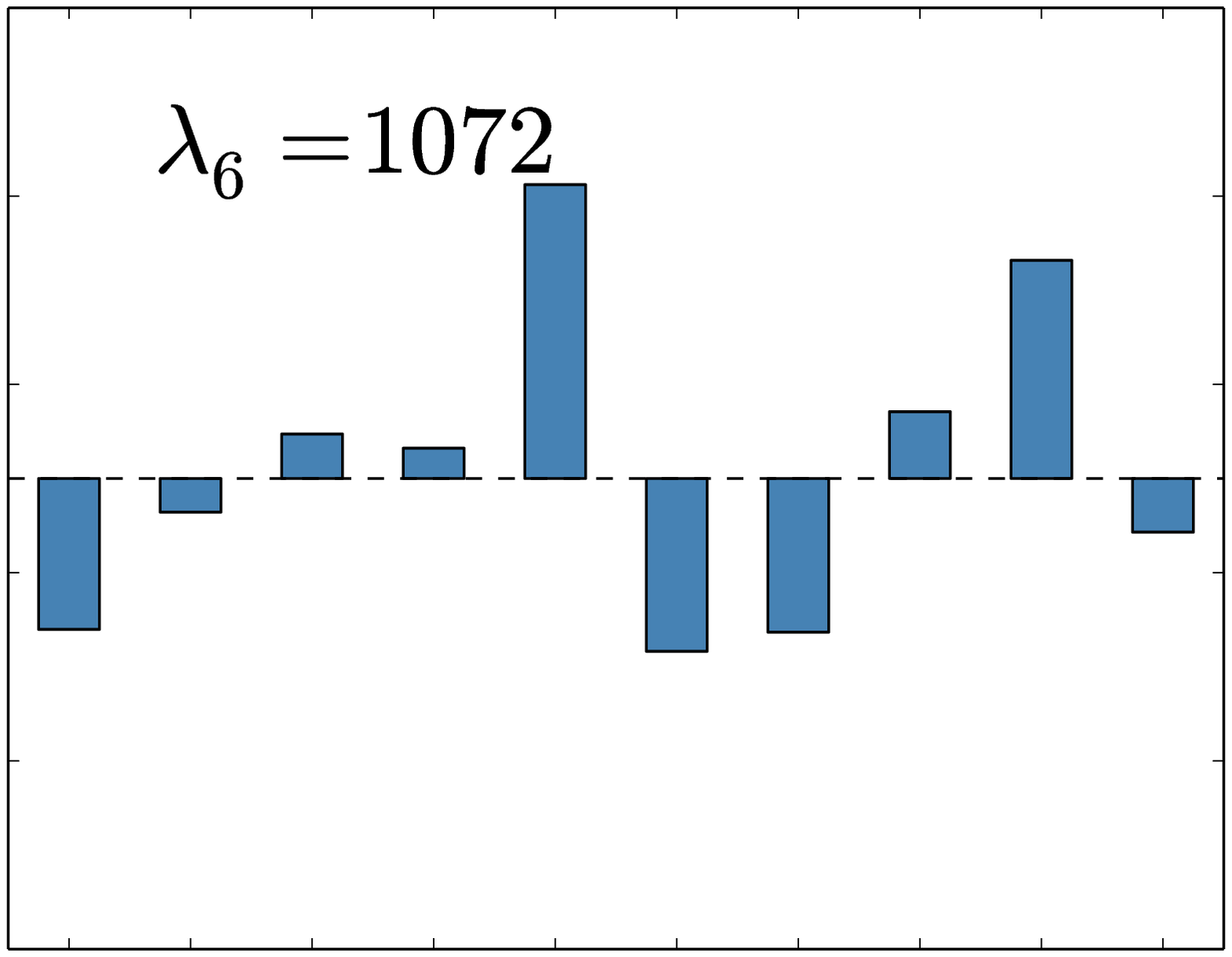} &
\includegraphics[scale=0.3,trim= 60 35 55 15,clip=true]{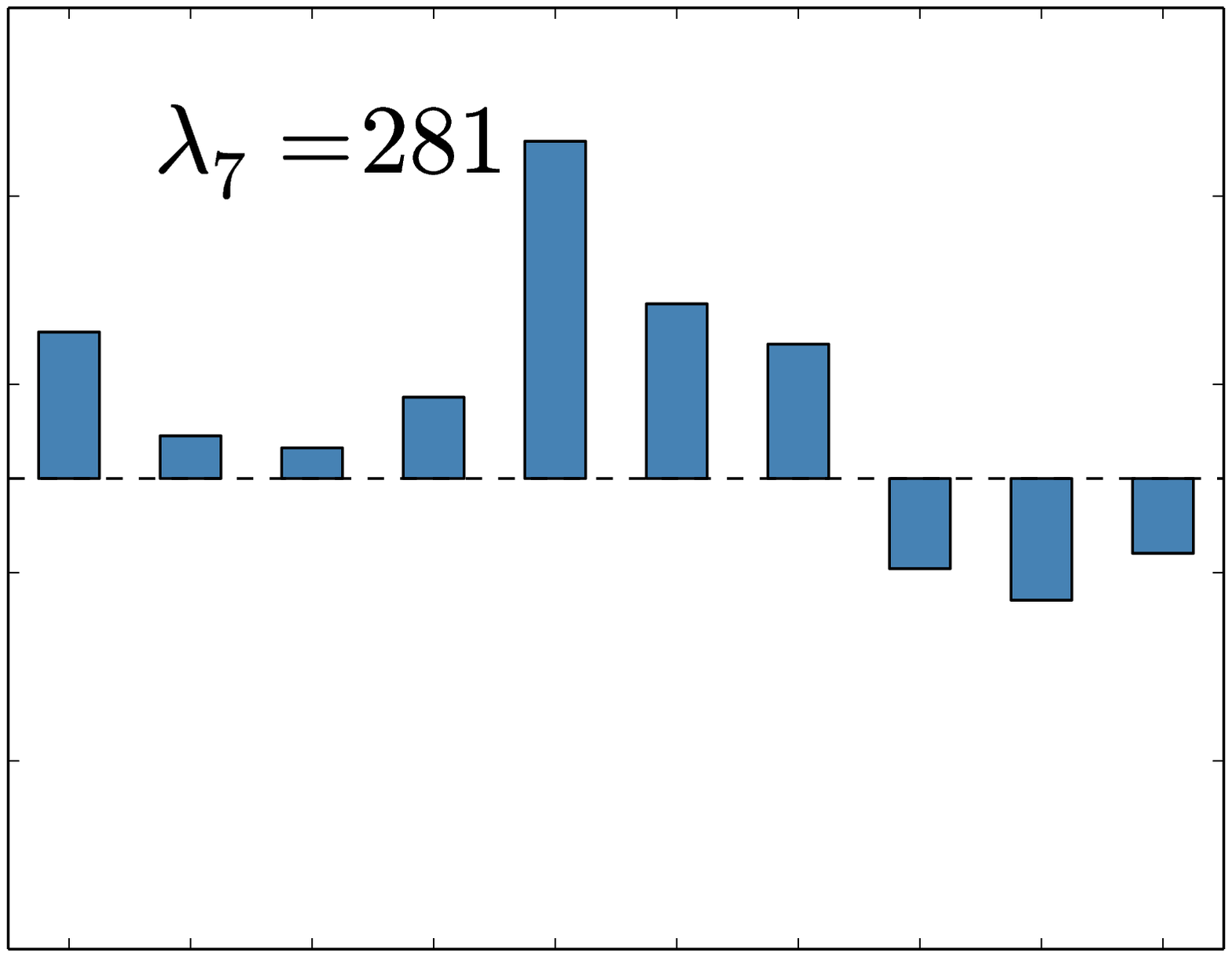}\\
\includegraphics[scale=0.3,trim= 5 0 55 15,clip=true]{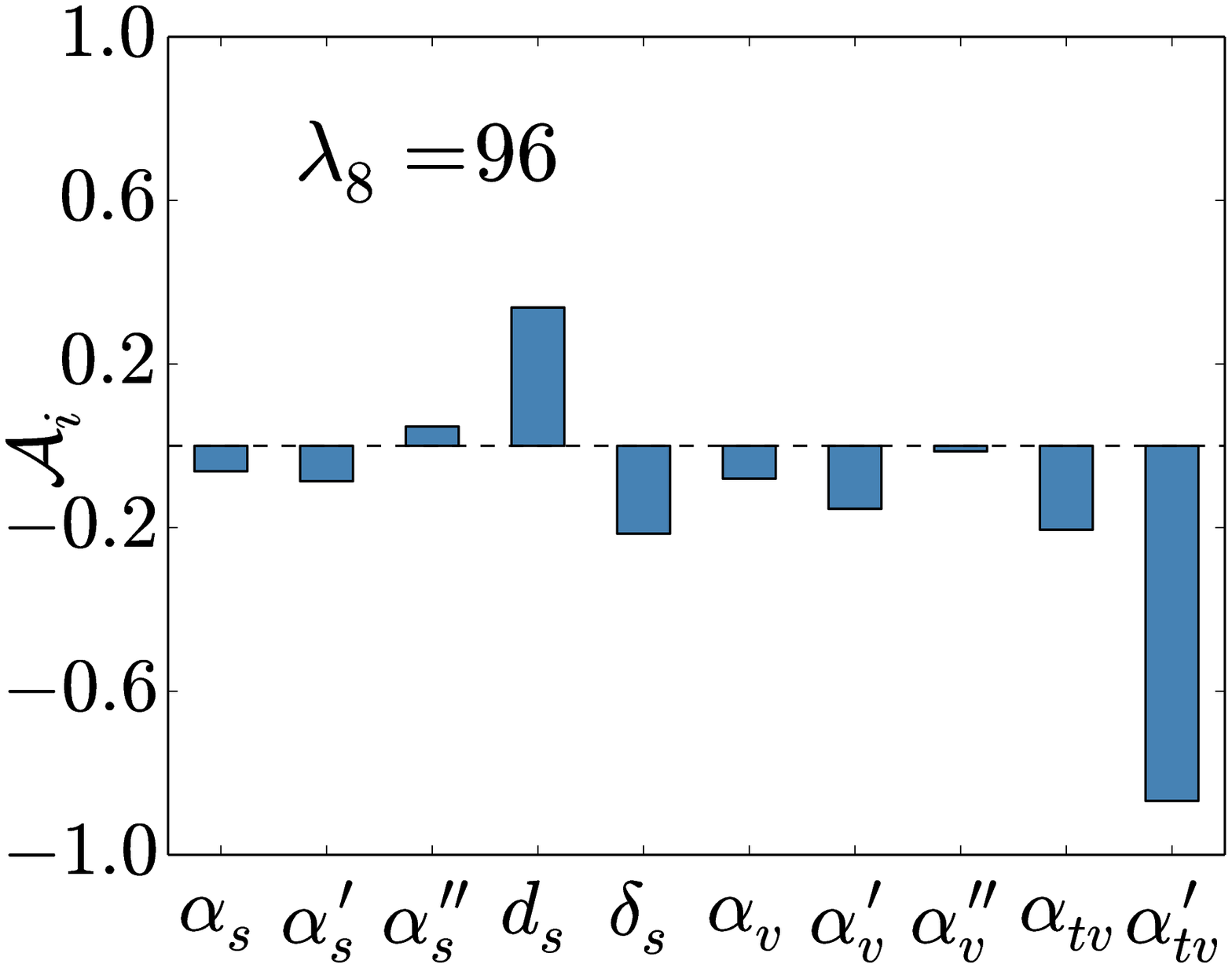} &
\includegraphics[scale=0.3,trim= 60 0 55 15,clip=true]{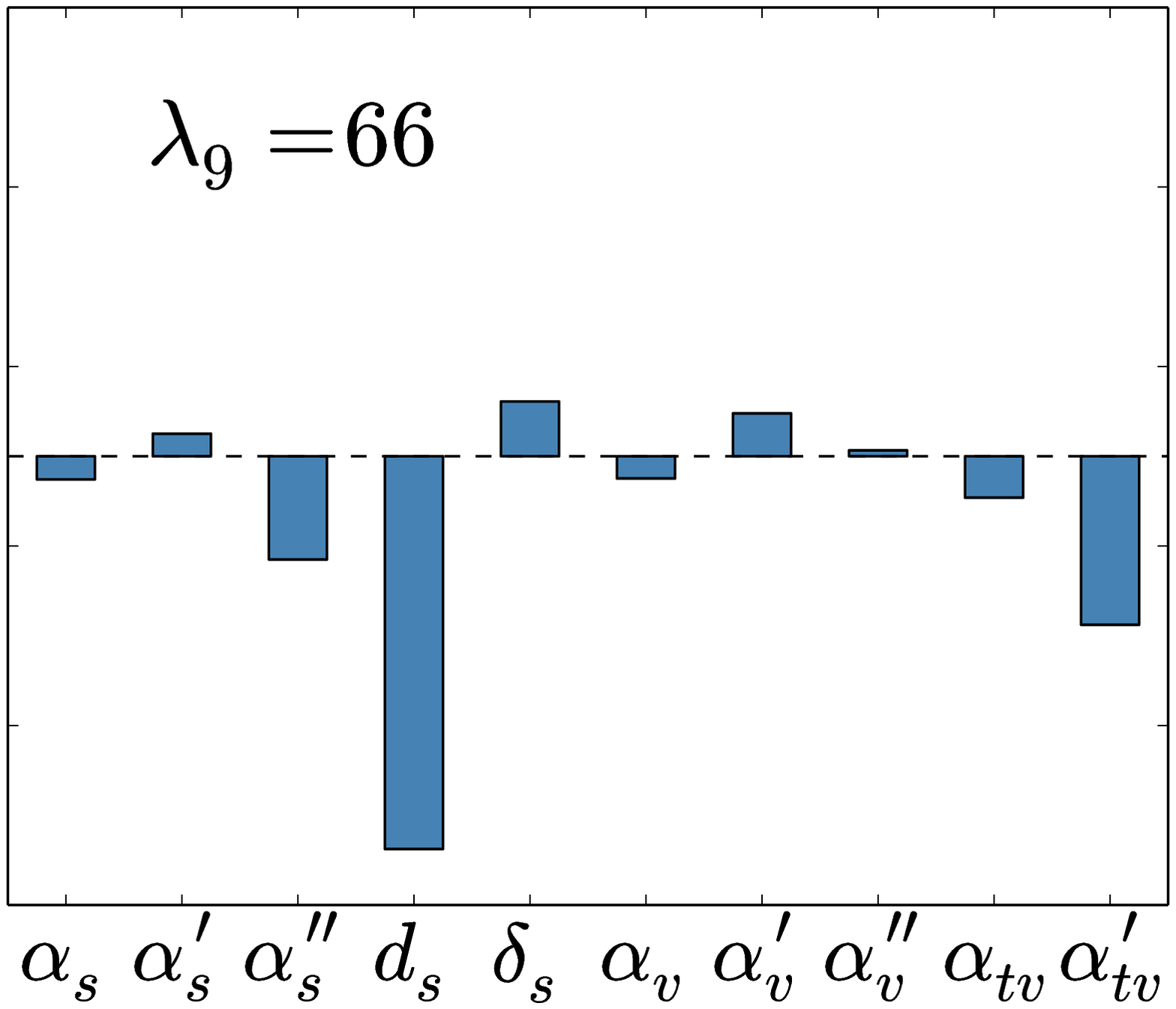}&
\includegraphics[scale=0.3,trim= 60 0 55 15,clip=true]{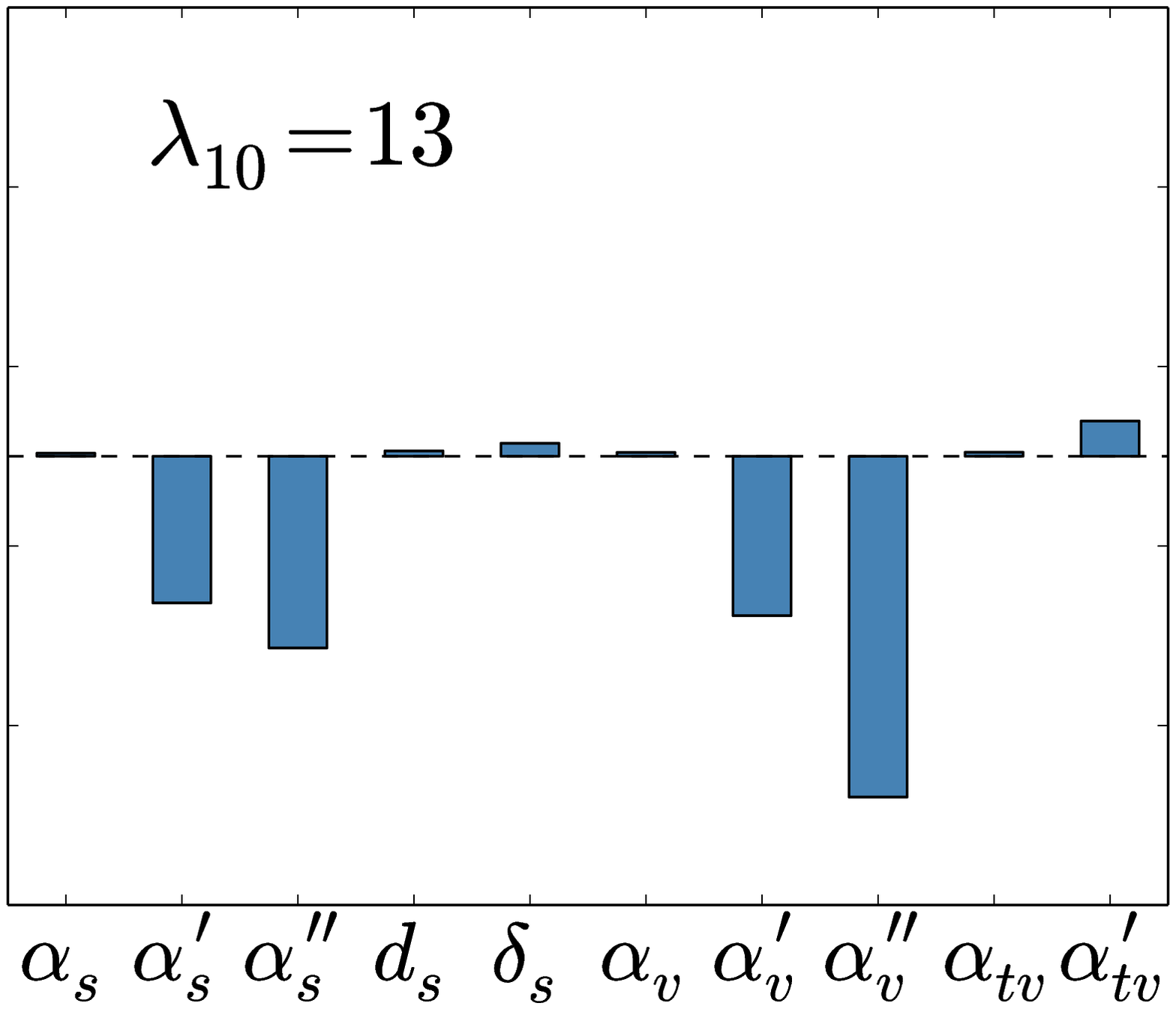}
\end{tabular}
\caption{\label{fig:modes-sinm} (Color online) Same as in the caption to 
Fig.~\ref{fig:mat-A}, but now the model parameter space includes  
 the strength $\delta_s$ of the derivative term in Eq.~(\ref{Lagrangian}), and 
 the surface energy of semi-infinite nuclear matter is added to the set 
 of pseudo-data used to calculate $\chi^2$.}
\end{figure}

To analyze the uncertainty and the corresponding correlations for the strength parameter 
$\delta_s$ of the derivative term in Eq.~(\ref{Lagrangian}), we have extended the 
calculation of the quality measure $\chi^2(\mathbf{p})$ 
and the corresponding matrix of second derivatives $\mathcal{M}$ in Eq.~(\ref{M}) to
include semi-infinite nuclear matter. For the surface energy 
we take the empirical value $a_s=17.5$ MeV with a 2\% uncertainty, the same 
as for the other pseudo-data. The dimension of the matrix $\mathcal{M}$
of second derivatives is now $10 \times 10$, and the diagonal 
matrix elements in order of decreasing values 
and the components of the corresponding eigenvectors are displayed in Fig.~\ref{fig:modes-sinm}. 
The distribution of components for most of the modes is 
very similar to those already shown in Fig.~\ref{fig:mat-A}, and we note that only 
the sixth and seventh eigenvectors contain significant amplitudes that arise from the derivative term. 
These eigenvectors also contain in-phase contributions from the isoscalar scalar and vector 
couplings, signaling that they are to some extent constrained by the effective mass.
Because of the contribution of the effective mass to the symmetry energy, 
modes number six and seven contain sizeable amplitudes from the isovector channel. 

\begin{figure}[htb]
\centering
\begin{tabular}{cc}
\includegraphics[scale=0.4]{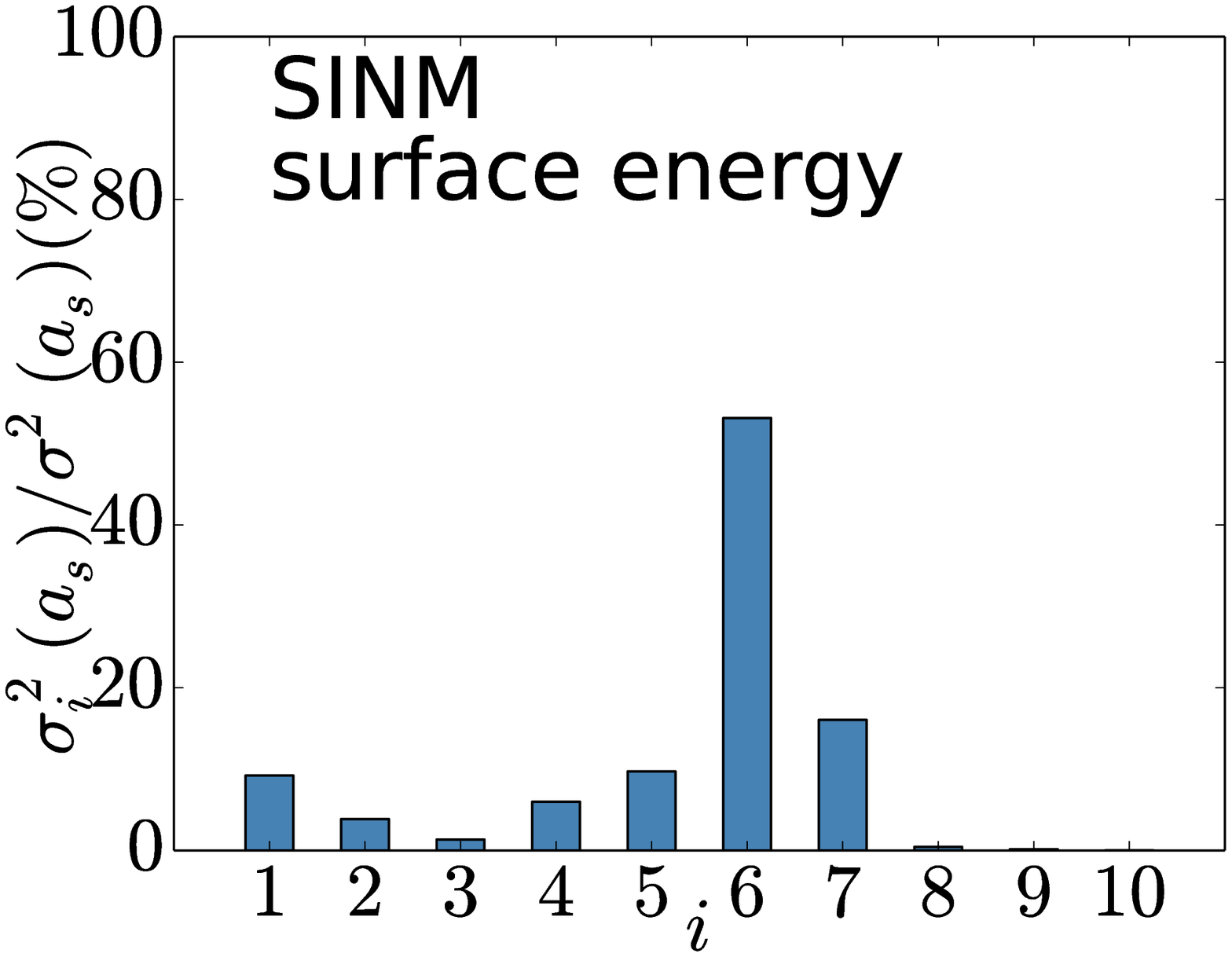} 
\includegraphics[scale=0.4]{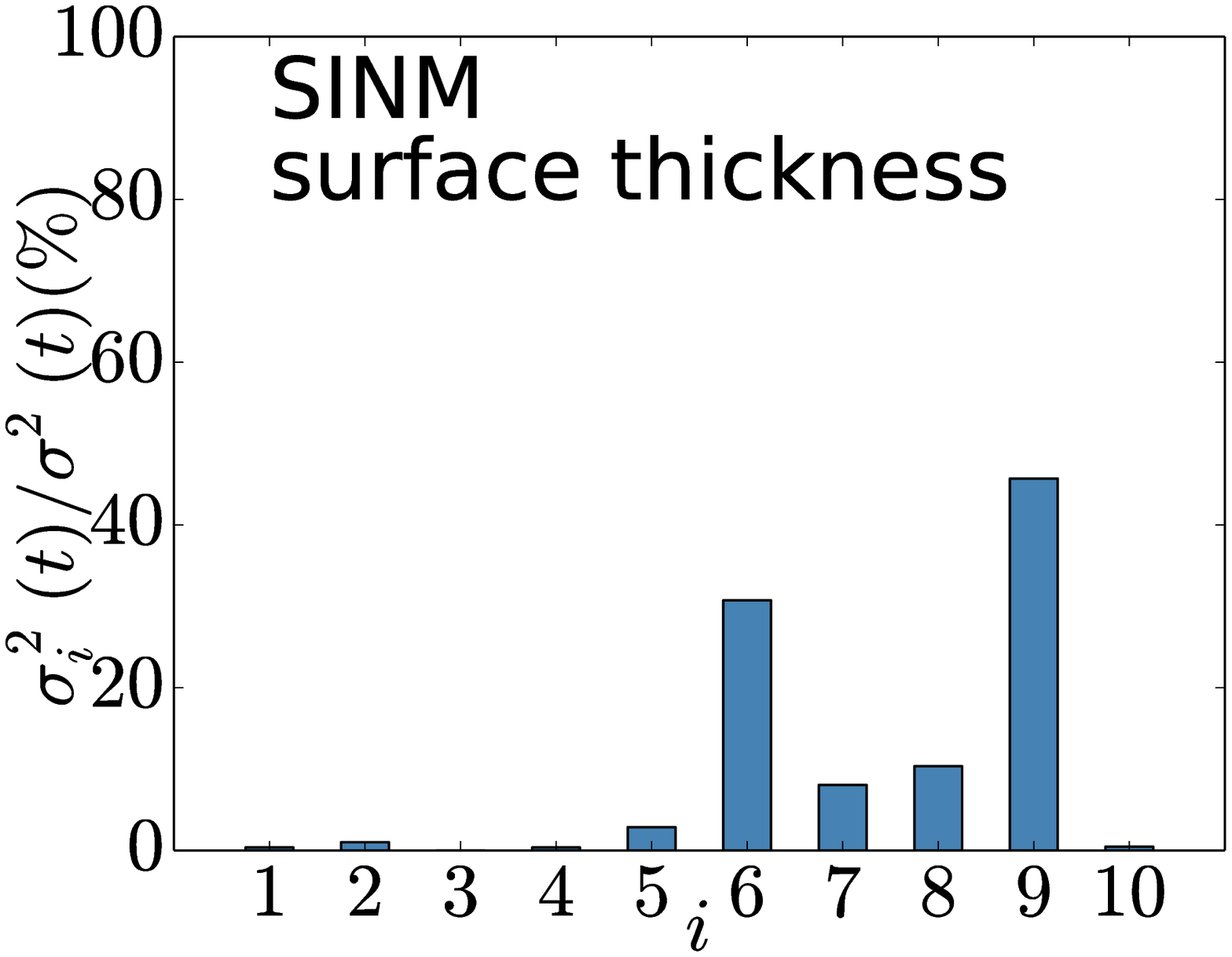} 
\end{tabular}
\caption{\label{fig:sigma-SINM} (Color online) Relative contributions in percentage 
of the ten linear combinations of model parameters that correspond to the eigenvectors of the 
matrix of second derivatives $\mathcal{M}$ in Eq.~(\ref{M}) 
(see Fig.~\ref{fig:modes-sinm}), to the variances of the surface energy and surface thickness of 
semi-infinite nuclear matter .}
\end{figure}

The theoretical uncertainties and correlation coefficients for the model parameters,
shown in the panels on the right of Fig.~\ref{fig:uncertainties-parameters} and 
and Fig.~\ref{fig:correlations-parameters}, respectively, 
are not significantly altered by the inclusion of the surface energy of semi-infinite nuclear matter 
in the set of pseudo-observables. 
We note, however, that the uncertainty of the parameter $\delta_s$ of the derivative term 
is larger than those of $\alpha_s(\rho_\mathrm{sat})$ and $\alpha_v(\rho_\mathrm{sat})$, 
with which $\delta_s$ displays significant correlation.

Finally, in Fig.~\ref{fig:sigma-SINM} we plot the relative contributions from the ten 
eigenvectors of the matrix $\mathcal{M}$ to the variance of the surface energy and surface thickness of 
semi-infinite nuclear matter (SINM).  The largest contribution to the variance of the
surface energy corresponds to mode six, which is dominated by the strength parameter 
of the isoscalar derivative term. The 
variance of the surface thickness, which has not been included in the set of pseudo-data 
used to calculate $\chi^2(\mathbf{p})$, displays an even more pronounced contribution from 
mode nine which predominantly corresponds to the parameter $d_s$ of the isoscalar scalar 
coupling (cf. Eq.~(\ref{parameters})).
\section{\label{secV} Finite nuclei}

\begin{figure}[htb]
\centering
\includegraphics[scale=0.65]{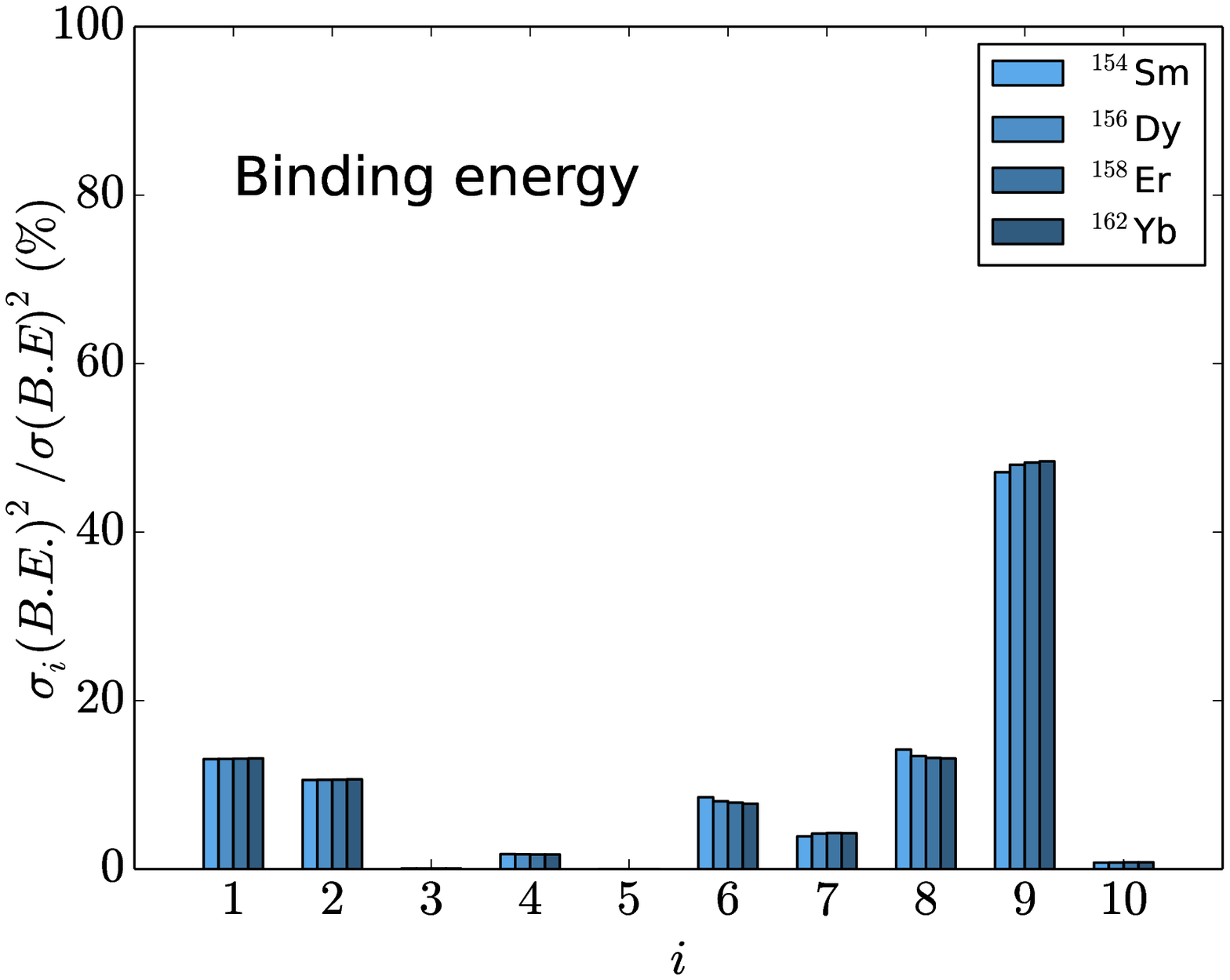}
\caption{\label{fig:be-rare-earth} (Color online)  Relative contributions in percentage 
of the ten linear combinations of model parameters that correspond to the eigenvectors of the 
matrix of second derivatives $\mathcal{M}$ Eq.~(\ref{M}) 
(see Fig.~\ref{fig:modes-sinm}), to the variances of the binding energy of rare-earth nuclei.}
\end{figure}
\begin{figure}[htb]
\centering
\includegraphics[scale=0.65]{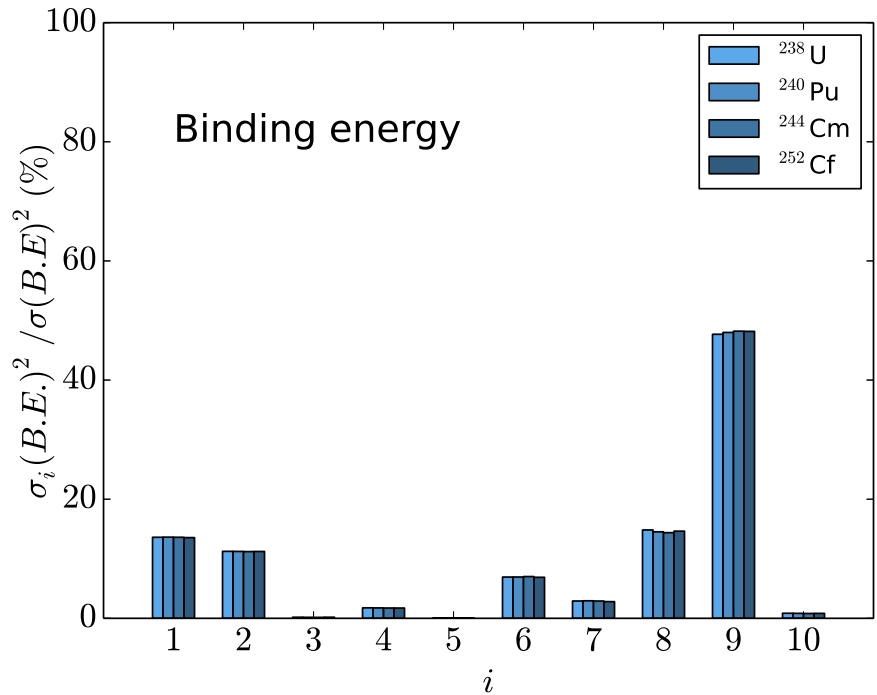}
\caption{\label{fig:be-actinides} (Color online) Same as in the caption to Fig.~\ref{fig:be-rare-earth} 
but  for the binding energy of actinide nuclei.  }
\end{figure}
\begin{figure}[htb]
\centering
\begin{tabular}{c}
\includegraphics[scale=0.65]{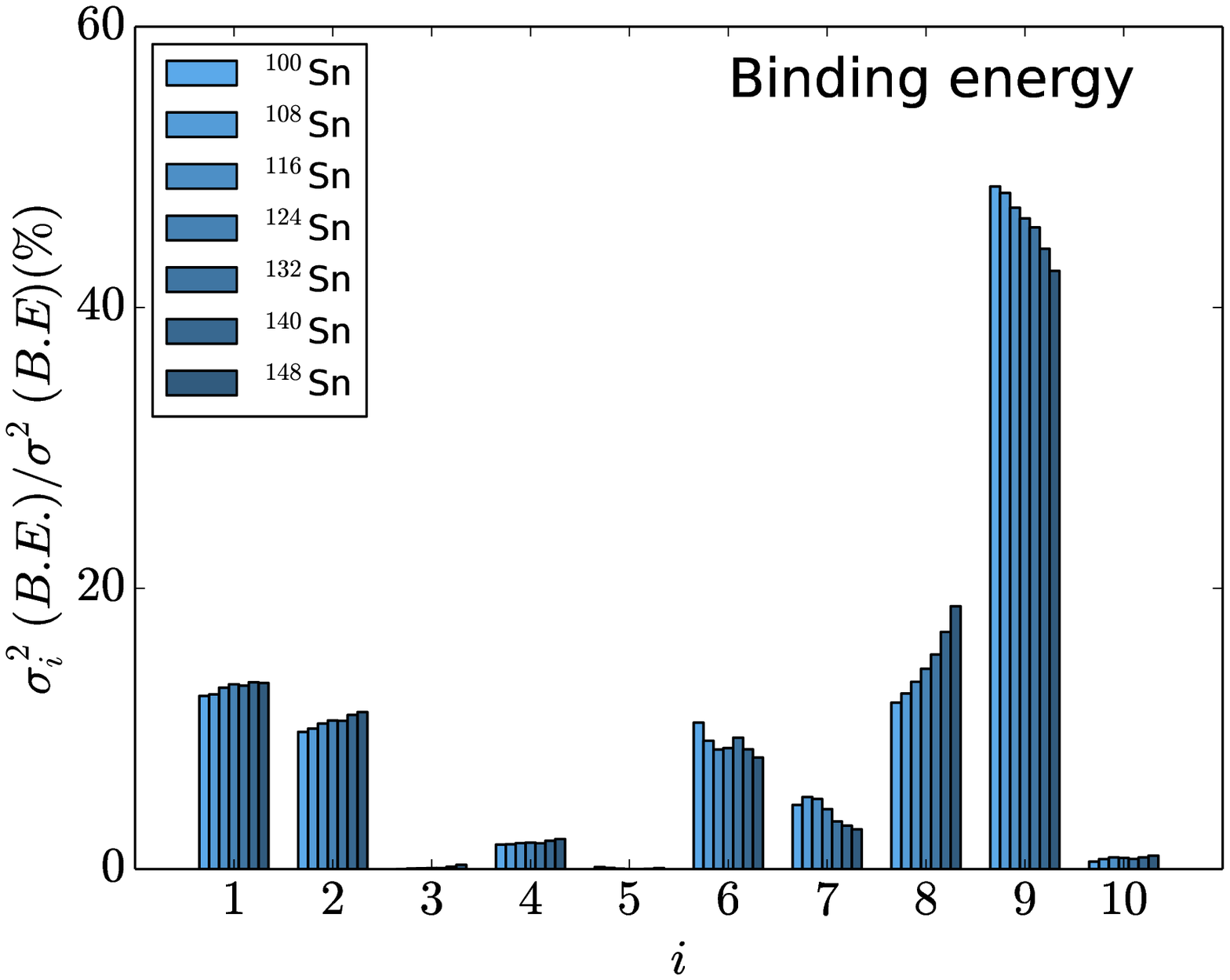} \\
\includegraphics[scale=0.65]{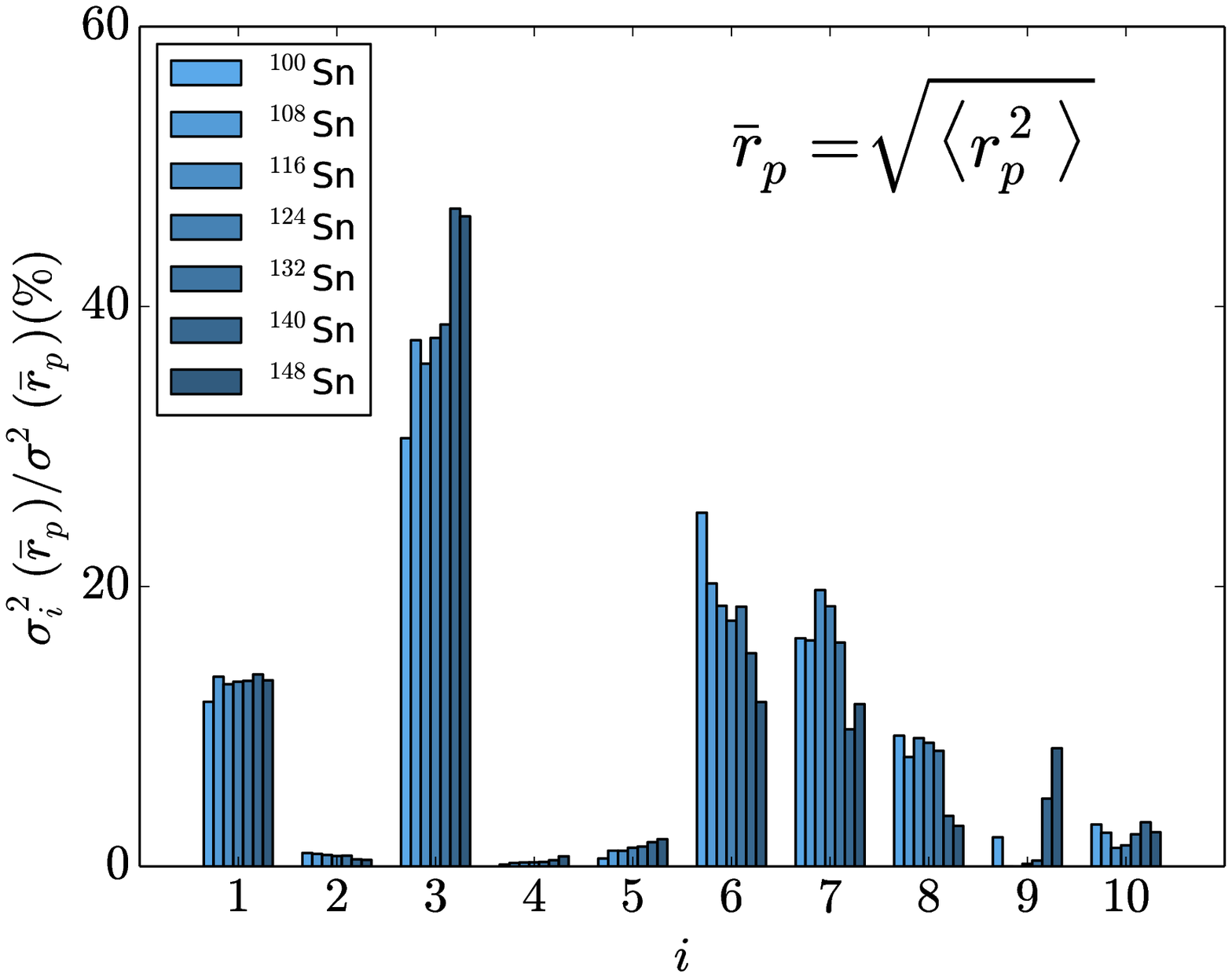}
\end{tabular}
\caption{\label{fig:be-tin} (Color online) Same as in the caption to 
Fig.~\ref{fig:be-rare-earth} but 
for the binding energies (upper panel), 
and radii of the proton distribution (lower panel) of tin isotopes.}
\end{figure}

In this section we extend the calculation of variances to ground-state 
observables of finite nuclei: binding energies and charge radii. Ground-state 
properties of spherical and deformed nuclei are computed in the framework of 
the relativistic Hartree-Bogoliubov model \cite{Vre05aR}.
In addition to the energy density functional in the particle-hole channel, 
pairing correlations are taken into account by using an interaction that is 
separable in momentum space, and determined by two parameters adjusted to 
reproduce the empirical pairing gap in symmetric nuclear matter. As already 
explained in Sec.~\ref{secIII}, the 10 parameters of the functional DD-PC1 
were determined by a fit to the experimental masses of 64 axially deformed nuclei
in the mass regions $A\approx 150-180$ and $A\approx 230-250$ \cite{Nik08}. 
In a series of subsequent studies it has been shown that this functional provides 
accurate predictions for binding energies, charge radii, deformation
parameters, neutron skin thickness, and excitation energies of giant
monopole and dipole resonances, as well as spectroscopic properties 
calculated using the generator coordinate method and/or the generalized 
quadrupole collective Hamiltonian \cite{Nik11}. In Figs.~\ref{fig:be-rare-earth} 
and \ref{fig:be-actinides} we display the relative contributions 
of the ten linear combinations of model parameters that correspond to the eigenvectors of the 
matrix of second derivatives $\mathcal{M}$ to the variances of the binding energy for 
a representative set of rare-earth and actinide nuclei, respectively. In both mass regions the 
largest contribution to the variance of the binding energy originates from the next-to-softest 
mode number nine, dominated by the parameter $d_s$ in the isoscalar-scalar 
channel (cf. Eq.~(\ref{parameters})). Modes number one and two correspond to 
out-of-phase contributions of the isoscalar 
couplings and their first derivatives at saturation density, respectively, 
and correspond to stiff directions in the 
parameter space. Finally, more pronounced contributions to the variance of 
the binding energy result also from the soft modes number six and eight, which  
correspond to combinations of isoscalar and isovector parameters: 
predominantly $\delta_s$ and $\alpha_{tv}$ in mode six, and $d_s$ and the 
derivative $\alpha^\prime_{tv}$ in mode eight. 

In contrast to rare-earth nuclei and actinides, spherically symmetric nuclei and, in particular, 
tin isotopes were not included in the adjustment of the parameters of DD-PC1. 
Nevertheless, one finds essentially the same distribution of eigenvectors 
of the matrix of second derivatives of $\chi^2(\mathbf{p})$ in the variance 
of binding energies of tin isotopes. This is shown in the upper panel of 
Fig.~\ref{fig:be-tin}, where we plot the relative contributions of the eigenmodes 
of $\mathcal{M}$ to the variance of the binding energies of Sn nuclei with mass 
number $100 \leq A \leq 148$. The fact that the largest contribution to the 
variances originates from the next-to-softest mode, that is, from a combination 
of parameters ($d_s$ and the derivative $\alpha^\prime_{tv}$) poorly constrained 
by the set of pseudo-data that determine $\chi^2(\mathbf{p})$, indicates that 
the choice for the ansatz of the density dependence of the isoscalar-scalar  
coupling (cf. Eq.~(\ref{parameters})) should be reexamined. We also note the 
increase of the relative contribution of the, predominantly isovector, mode eight 
in tin isotopes with a larger neutron excess. It is particularly interesting to compare the 
relative contributions to the variance of the binding energy to those of an 
observable that was not included in the fit of the parameters of DD-PC1. 
The lower panel  of Fig.~\ref{fig:be-tin} displays the relative contributions 
of the eigenmodes of $\mathcal{M}$ to the variance of the radius of the 
proton distribution of tin isotopes. In this case the variance for all isotopes 
is dominated by the relatively stiff combination of parameters that 
corresponds to mode three, and its relative contribution increases with 
neutron number. The significant contributions from the soft modes six, seven, and 
eight decrease in neutron-rich tin isotopes whereas, as one would expect, the 
out-of-phase contributions of isoscalar couplings in the stiff mode one do not 
show significant variation with neutron number.

\section{\label{secVI}Conclusion}
Nuclear density functional theory (NDFT) provides a unified framework for studies of ground-state 
properties and collective excitations across the nuclide chart. Even though methods and structure 
models based on DFT have been extremely successful in analyzing a variety of nuclear 
properties and predicting new structure phenomena, a fully microscopic 
foundation of nuclear energy density functionals (NEDFs), based on and constrained by the 
underlying theory of strong interactions, has yet to be established. However, even if this task 
is accomplished in future, the parameters of a NEDF will have to be fine-tuned to data on 
finite, medium-heavy and heavy nuclei. This is because of the inherent complexity of the 
effective in-medium inter-nucleon interactions that cannot fully be unfolded starting from 
the fundamental low-energy degrees of freedom, nor by data on nucleon-nucleon scattering 
and few-nucleon systems. 

Some of the most successful NEDFs are semi-phenomenological and approximate 
the exact unknown functional by an expansion in powers of ground-state nucleon densities and 
currents and their gradients, and/or assume a relatively simple ansatz for the density 
dependence of the effective inter-nucleon interactions, often based on a microscopic nuclear 
matter equation of state. The problem than becomes how to select the most efficient
functional form and/or parametrization of the density dependence, considering the fact that 
data on ground-state properties can only constrain a very limited set of terms and parameters 
in a general expansion of the nuclear EDF. Until recently the standard procedure of adjusting
nuclear density functionals was to perform a least-squares fit of parameters simultaneously
to empirical properties of symmetric and asymmetric nuclear matter, and to selected 
ground-state data of a small set of spherical closed-shell nuclei. A new generation of 
density functionals is currently being developed that, on the one hand, is more firmly constrained 
by microscopic treatments of effective inter-nucleon interactions and, on the other hand, their 
parameters are adjusted to much larger data sets of ground-state properties, including both 
spherical and deformed nuclei. Methods of statistical analysis can be used to assess the 
uniqueness and predictive power of particular functionals, as well as the stability or 
sensitivity of model parameters. These methods can also be used to determine the type 
of data that better constrain model parameters. 

In this work we have analyzed a particular class of relativistic energy density functionals 
characterized by contact (point-coupling) effective nucleon-nucleon interactions and 
density-dependent coupling parameters. The ``best-fit model'', the functional DD-PC1, was 
adjusted in a multistep parameter fit to experimental masses of a large set of deformed 
heavy nuclei. We have used covariance analysis to examine the stability of this 
functional in nuclear matter, and to determine weakly and strongly constrained combinations 
of parameters. In particular, instead of analyzing uncertainties and correlations between 
the individual parameters of the given functional, we have examined correlations between 
the lowest-order terms in a Taylor expansion of the density-dependent coupling parameters   
around the saturation point in nuclear matter. To this end, we have produced a set of pseudo-observables 
in infinite and semi-infinite nuclear matter, computed with the
functional DD-PC1, and used these data to compose a quality measure $\chi^2$.
In the spirit of statistical analysis, we have analyzed the behavior of  $\chi^2$ around the minimum. In particular, 
we have computed uncertainties of model parameters and correlation coefficients 
between parameters, as well as the eigenvectors and eigenvalues of the 
matrix of second derivatives of $\chi^2$. This has allowed to determine stiff and soft 
directions in the parameter space, that is, to deduce which combinations of model 
parameters are firmly constrained by nuclear matter pseudo-data, and which 
combinations are poorly determined by the quality measure $\chi^2$. 
In addition, we have also analyzed the uncertainties of observables that were not included 
in the calculation of $\chi^2$: binding energy of asymmetric nuclear matter, surface thickness 
of semi-infinite nuclear matter, binding energies and charge radii of finite nuclei. 
The present covariance analysis has shown that, even though the functional DD-PC1 has 
been successfully employed in a series of spectroscopic studies and used to 
predict various nuclear properties, several combinations of model parameters 
appear to be weakly constrained in nuclear matter and some of them produce rather large 
uncertainties for observables of finite nuclei. The results of this analysis also show that 
the adopted ansatz for the density dependence of the coupling parameters of DD-PC1 
should be reexamined. 

More generally, the results obtained in this work illustrate how a simple analysis of 
the quality measure $\chi^2$ around the minimum in nuclear matter can be used as 
a starting point in the determination of the functional density dependence of a nuclear EDF, 
and in the selection of the type of data that can more firmly constrain the values of model parameters.  
\clearpage
\section*{References}
\bibliographystyle{unsrt}
\bibliography{Covariances}

\end{document}